\documentclass[letter,12pt]{article} 

\usepackage[latin1]{inputenc} 
\usepackage{amssymb}
\usepackage{cite}
\usepackage{amsfonts}
\usepackage{amsthm}
\usepackage{multicol}
\usepackage{multirow}
\usepackage{amsmath}
\usepackage{graphicx}
\usepackage{float}
\usepackage{hyperref}

\begin{document}

\begin{center}
\textbf{Influence of pairing and deformation on charge exchange transitions} 

\vspace{0.25cm}

\textbf{A. Carranza M. $^{(1)}$, S. Pittel $^{(2)}$, Jorge G. Hirsch $^{(1)}$.} 

(1) Instituto de Ciencias Nucleares, Universidad Nacional Aut\'onoma de M\'exico, 04510, M\'exico CDMX,
M\'exico.

(2) Bartol Research Institute and Department of Physics and Astronomy, University of Delaware,
Newark, Delaware 19716, USA.
\end{center}

\section*{Abstract}

 We describe the importance of charge-exchange reactions, and in particular Gamow-Teller transitions, first to astrophysical processes and double beta decay, and then to the understanding of nuclear structure.  In our review of their role in nuclear structure we first provide an overview of some of the key steps in the emergence of our current understanding of the structure of nuclei,  including the central role played by the isovector pairing and the quadrupole-quadrupole channels in the description of  energy spectra and in the manifestation of collective modes, some associated with deformation of the nuclear shape. We then turned our focus to Gamow-Teller (GT) transitions in relatively light nuclei, especially in the $2p1f$ shell, where isoscalar pairing may be playing a role in competition with the isovector pairing that dominates in heavier regions. Following a summary of the progress made in recent years on this subject, we report a systematic shell model study aimed at providing further clarification as to how these pairing modes compete. In this study, we use a schematic Hamiltonian that contains a quadrupole-quadrupole interaction as well as both isoscalar and isovector pairing interactions. We first find an optimal set of Hamiltonian parameters for the model, to provide a starting point from which to vary the relevant pairing strengths and thus assess how this impacts the behavior of GT transitions and the corresponding energy spectra and rotational properties of the various  nuclei involved in the decays. The analysis includes as an important theme a comparison with experimental data. The need to suppress the isoscalar pairing mode when treating nuclei with a neutron excess to avoid producing spurious results for the ground state spin and parity with the simplified Hamiltonian is highlighted. Varying the strength parameters for the two pairing modes is found to exhibit different but systematic effects on GT transition properties and on the corresponding energy spectra, which are detailed.

\section{Introduction}

Gamow-Teller (GT) transitions provide an important and useful tool in the exploration of nuclei \cite{Caurier1995,Cole2012,Kumar2016,Kumar2016b}. They play a key role in the $\beta$ decay and electron capture processes that arise in stellar evolution \cite{Langanke2003,Balasi2015}, in the double beta decay process \cite{Saakyan2013,Dolinsky2019} and in neutrino nucleosynthesis \cite{Heger2005,Suzuki2013,Byelikov2007}. Furthermore, they are very useful for the testing of nuclear models.

There are two types of GT transitions;  $ GT_{+}$ in which a proton is changed into a neutron, and $ GT_{-}$ in which a neutron is changed into a proton. The transition strength B(GT) can be obtained from $\beta$ decay studies, but with excitation energies limited by the $Q$ values of the decays. On the other hand, with charge-exchange (CE) reactions, such as $(p,n)$, $(n,p)$, $ (d,^{2}He)$, or $ (^{3}He,t) $, it is possible to access GT transitions for large values in energy without the $Q$ value restriction. Experimental measurements for such reactions at angles close to $ 0^{\circ} $ and with an incident energy above $100~MeV$/nucleon provide valuable information about GT transitions.

In this work, we describe the relevance of charge-exchange reactions, and in particular Gamow-Teller transitions, in the various areas noted above in which it is known to play an important role. We begin with a description of its role in nucleosynthesis in astrophysical environments and in double beta decay, a rare process whose neutrinoless mode, if observed, would shed light on the understanding of neutrino properties. While some experimental results are mentioned, the emphasis here is on the theoretical developments in the last few decades.

We then turn to the important role of Gamow Teller transitions in nuclear structure. The dominance of the quadrupole-quadrupole interaction in the particle-hole channel and isovector (J=0) pairing in the particle-particle channel serves as the  microscopic foundation for the use of the pairing plus quadrupole Hamiltonian, whose development is briefly reviewed.

For a meaningful description of relatively light nuclei with $N\approx Z$, however, it is important to also consider the possible importance of proton-neutron (pn) pairing. Proton-neutron pairing can arise in two channels, isoscalar (T=0) and isovector (T=1), in both of which the neutron and proton can have net zero orbital angular momentum and thus exploit the short-range nuclear force.  Its role in relatively light nuclei has been studied in recent years  with a variety of formalisms. Some of the earliest work considered the extension of mean field techniques like  BCS or Hartree-Fock-Bogolyubov (HFB) \cite{Goodman1979, Doba04} to include the proton-neutron pairing channel. Its possible importance in $N \approx Z$ nuclei has also been studied recently in the context of the nuclear shell model in several recent works \cite{Petermann2007,Lei2011}. Of particular interest is the isoscalar pairing channel as it is expected to be especially important for nuclei with $N=Z$ and $ N \approx Z $ \cite{Lerma2007,Sandulescu2009}.

We focus our attention on GT transition intensities in the $2p1f$ shell and particularly nuclei in the $A=40-48$ region. For these nuclei, the transitions $GT_{-}$ and the reactions $(p,n)$ are the primary experimental tools used to obtain the distribution of intensities, but the reaction $ (^{3}He,t) $ has served as an alternative experimental tool. This review includes many of the most cited published works, but we apologize in advance for any omissions.

Though as noted earlier, mean field techniques have often been used to treat the interplay of the various pairing modes in nuclei near $N=Z$, it is known that they can lead to serious errors because of their violation of symmetries \cite{Dobes1998}. [Note: Recent efforts \cite{Dobaczewski} to build a symmetry-restored mean field approach have proven promising.]
For this reason it is especially useful to study how the various modes of pairing compete in nuclear systems near $N=Z$  in the context of the nuclear shell model, whereby it is possible to treat all pairing modes on an equal footing,  to preserve all symmetries,  and also to naturally incorporate the effects of deformation which are critical in the regions in which these pairing modes are thought to be important.

For the above reasons, we have chosen to use the shell model to assess how the various pairing modes affect Gamow-Teller intensities in light $N \approx Z$ nuclei. Our approach is inspired by earlier work \cite{Lei2011} where a simple parametrized shell model Hamiltonian was used in order to have a a tool for systematically isolating the effects of the two pairing channels. The model includes a quadrupole-quadrupole interaction as well as both isoscalar and isovector pairing interactions. By leaving out other components of realistic nuclear Hamiltonians, we can focus directly on the effects of the different pairing modes, albeit at the cost of missing some important features of the structure of the nuclei we consider.

We consider in this analysis even-mass nuclei from A=42-48. We focus
on the fragmentation of GT transition strengths, but also consider the energy spectra of both the parent and daughter nuclei involved.

A summary of the paper is as follows. In Sec. II we review the role of Gamow-Teller transitions in astrophysics and particle physics, briefly describe the pairing and quadrupole-quadrupole interactions and then summarize earlier shell model investigations of Gamow-Teller transitions in the $2p1f$ shell.  In Sec. III we briefly describe the model we use and in Sec. IV describe selected results of our analysis for GT transition strengths. We leave a discussion of the corresponding results for energy properties to the Appendix, where we also show results of a related Hamiltonian \cite{Lei2011} differing only in the choice of single-particle energies.

\section{The relevance of Gamow-Teller transitions}

\subsection{Charge exchange reactions and the Gamow-Teller strength function}

Charge exchange reactions employing nuclear projectiles and targets play a fundamental role in the study of the isospin- and spin-isospin dependent response of nuclei. They reveal important aspects of  nuclear dynamics, connecting strong and weak interactions \cite{Lenske2019}. One of them is the Gamow-Teller (GT) strength distribution,  which is usually inaccessible to beta decays, but can be obtained with high resolution using reactions such as (p, n) (i.e., isospin-lowering) and (n, p) (i.e., isospin-raising), performed at intermediate energies and at nearly zero momentum transfer \cite{Ichimurai2006,Frekers2018}. There is a proportionality between single charge-exchange reaction cross sections in the forward direction, e.g. (p, n) and ($^3$He,t),  and the Gamow-Teller (GT) strength into the same final nuclear states. In nuclei with a neutron excess, the main contribution to the GT strengths comes from the removal of a neutron from an occupied single-particle state and the placement of a proton into an unoccupied state having either the same  quantum numbers or those of the spin-orbit partner. However, the opposite channel, explored with (n, p) and (d,$^2$He) reactions, is Pauli forbidden in medium-heavy nuclei and can only be effective if the Fermi surface is smeared out, which can introduce a radial dependence that is usually not included in the analysis \cite{Amos2007}.

The smearing of the Fermi surface is in many cases obtained introducing like-particle isovector (J=0) pairing correlations at mean field level through the BCS approach, while more sophisticated deformed selfconsistent Hartree-Fock calculations with density-dependent Skyrme forces are also employed \cite{Sarriguren2001,Moreno2006}. Residual proton-neutron  interactions in the particle-hole and particle-particle channels are introduced employing the proton-neutron Quasiparticle Random Phase Approximation (pnQRPA) \cite{Cha83,Nabi2013,Niu2016,Sarriguren2018}. This interaction gives rise to collective spin modes such as the giant Gamow-Teller resonance \cite{Osterfeld1992}.

It has been observed that the measured total Gamow-Teller transition strength in the resonance region is much less than a model-independent sum rule predicts, the so called "quenching" phenomena \cite{Osterfeld1991,Suhonen2013,Suhonen2017}. It has been partially explained introducing coupling with $\Delta$-hole excitations \cite{Hirsch1990,Udagawa1994,Cattapan2002}, associated with nuclear correlations,  which cause many of the resulting peaks to be weak enough to become unobservable in full shell model calculations \cite{Caurier1995,Caurier1995b,Gysbers2019}, to a Schiff part of the GT strength to higher energies due to  2-particle-2-hole excitations\cite{Hirsch1988}, tensor correlations \cite{Bai2009}, finite momentum transfer in relativistic QRPA descriptions \cite{Marketin2012}, and with the presence of two- and three-body weak currents \cite{Wang2018b,Gysbers2019}.

The theoretical estimate of GT nuclear matrix elements of beta-decay and electron-capture processes in heavy nuclei, which are usually deformed, has seen important developments in the last decades.
One of the latest involves employing the projected shell model with many multi-quasiparticle configurations included in the basis, which shows that the B(GT) distributions can have a strong dependence on the detailed microscopic structure of the relevant states of both the parent and daughter nuclei \cite{Wang2018}.

\subsection{Charge-exchange reactions in astrophysics}

Many relevant astrophysical processes like stellar burning, neutrino nucleosynthesis, explosive hydrogen burning and core-collapse supernovae involve nuclear weak interactions \cite{Rolfs1988}. Their description requires state of the art nuclear models associated with experimental data coming from radioactive ion-beam facilities \cite{Langanke2003,Bertulani2016}. Supernova explosions are associated with the collapse of the core of a massive star. The dynamics of this process involves electron capture on thousands of nuclei, whose rates must be estimated employing microscopic models \cite{Juodagalvis2010}.

Neutrinos generated in supernova explosions induce nuclear reactions which play an important role in the nucleosynthesis of heavy elements \cite{Ejiri2000,Ejiri2019}.
Recent advances in their description have benefited from new experimental data on allowed Gamow-Teller strength distributions, which are accurately reproduced employing improved nuclear models and computer hardware capabilities, including shell model diagonalization and the proton-neutron Quasiparticle Random Phase Approximation (pnQRPA) \cite {Balasi2015}. As most nuclei relevant to astrophysical processes have a large neutron excess, with proton and neutrons occupying different nuclear major shells, partial occupation numbers must be estimated. In the stellar high density environments, finite temperature dependent occupations are estimated employing the shell model Monte Carlo approach, later combined with the RPA to unblock the Gamow-Teller transitions at all temperatures relevant to core-collapse supernovae \cite{Langanke2001}.

\subsection{The double beta decay}

One of the nuclear processes associated with the weak interaction that continues to attract much experimental and theoretical attention is double beta decay \cite{Tomoda1991,Faessler1998,Suhonen1998,Vergados2002,Zdesenko2003,Elliot2004,Avignone2008,Barabash2011,Vergados2012,Bilenky2015,Engel2017}.
Two-neutrino double-beta decay has the longest radioactive half-lives ever observed, an outstanding experimental achievement that continues with new developments \cite{Saakyan2013}.
Its neutrinoless mode  is a forbidden, lepton-number-violating nuclear transition whose observation would have fundamental implications for neutrino physics and cosmology. A wide range of experiments has been performed and are in execution or planned to discover this decay, unseen up to now \cite{Dolinsky2019}. The theoretical estimate of the decay rates for the two-neutrino and neutrinoless modes are defined as second-order perturbative expressions starting from an effective electroweak Lagrangian \cite{Suhonen1998}. The neutrinoless mode can proceed through mechanisms involving light Majorana neutrinos, heavy Majorana neutrinos, sterile neutrinos and Majorons.

Theoretical calculations involve the evaluation of nuclear transition matrix elements, which have been estimated using a variety of nuclear models.
{\em Shell model calculations} have been used to estimate the double-beta decay nuclear matrix elements for $^{48}$Ca \cite{Zhao1990,Caurier1990,Retamosa1995,Horoi2007,Horoi2010} and in heavier nuclei\cite{Caurier1996,Horoi2013}.
Large-scale shell-model calculations for $^{48}$Ca, including two harmonic oscillator shells ($2s1d$ and $2p1f$) found that the neutrinoless double-beta decay nuclear matrix element is enhanced by about 30\% compared to $2p1f$-shell calculations \cite{Iwata2016}. With this formalism a very good linear correlation between the double Gamow-Teller transition to the ground state of the final nucleus and the neutrinoless double beta decay matrix element has been observed \cite{Shimizu2018}. Shell model Monte Carlo methods also provide valuable results \cite{Koonin1997}.
The {\em Quasiparticle Random Phase Approximation} (QRPA) provides a consistent treatment of both particle-hole and particle-particle interactions in calculations for the nuclear matrix elements governing two-neutrino and neutrinoless double-beta decay \cite{Vogel1986}. It opened the way to resolve the discrepancy between experimental and calculated two-neutrino decay rates, employing both schematic  \cite{Engel1988,Muto1988,Hirsch1990} and realistic interactions \cite{Civitarese1987,Staudt1990,Pirinen2015}. The suppression was also found employing a generalized-seniority-based truncation scheme \cite{Engel1989}, and in QRPA calculations including particle number projection \cite{Civitarese1990}. Extensions of these techniques have been widely applied to estimate the neutrinoless nuclear matrix elements \cite{Suhonen1991,Suhonen2011,Suhonen2012,Simkovic2013}, the decay to excited states \cite{Griffiths1992,Suhonen1993,Suhonen1994,Civitarese1994,Aunola1996}, and the influence of proton-neutron pairing \cite{Cheuon1993}. To avoid divergences found in the calculations, a renormalized  QRPA version was introduced \cite{Toivanen1995,Schwieger1996,Toivanen1997} which had its own difficulties\cite{Hirsch1996,Engel1997,Hirsch1997,Stoica2001,Rodin2003,Rodin2006}.

The influence of deformation on double beta decay rates was studied with the deformed QRPA formalism \cite{Raduta1993,Simkovic2004,Alvarez2004,Yousef2008} and the Pseudo SU(3) scheme\cite{Castanos1994,Hirsch1994}. The projected-Hartree-Fock-Bogoliubov (PHFB) approach, employing a pairing plus multipole type of effective two-body interaction, shows that deformation plays a crucial role in the nuclear structure aspects of the decays \cite{Singh2007,Chaturvedi2008,Rath2010,Rath2013,Rath2019}.

The NUMEN project employs an innovative technique to access the nuclear matrix elements entering the expression of the lifetime of the double beta decay by cross section measurements of heavy-ion induced Double Charge Exchange (DCE) reactions.  It has reached the experimental resolution and sensitivity required for an accurate measurement of the DCE cross sections at forward angles. However, the tiny values of such cross sections and the resolution requirements demand beam intensities much larger than those manageable with the present facility \cite{NUMEN2018}.

\subsection{The quadrupole-quadrupole and pairing Hamiltonians}

As needed background for our discussion of the role of Gamow Teller transitions in nuclear structure physics, we first give a brief reminder of some important concepts and early developments in nuclear structure theory.

 Isovector pairing interactions, those coupling like particles to zero total angular momentum \cite{Dean2003}, play a fundamental role in low energy nuclear structure, and are particularly relevant in the understanding of mass differences between even and odd-mass nuclei. Extending the Bardeen, Cooper and Schrieffer (BCS) description of superconductivity \cite{BCS1957} as a number-nonconserving state of coherent pairs, Bohr, Mottelson and Pines \cite{Bohr1958} proposed a similar physics mechanism to explain the large gaps seen in the spectra of even-even atomic nuclei, later corrected for finite size effects employing particle number projection \cite{Kerman1961,Nogami1964,Dietrich1964}. The exact solution of the isovector pairing Hamiltonian was presented by Richardson and Sherman in 1963 \cite{Richardson1963}. In the last decades it was extended to families of exactly-solvable models, called generically Richardson-Gaudin (RG) models\cite{Dukelsky2004,Dukelsky2013}, which found application in different areas of quantum many-body physics including mesoscopic systems, condensed matter, quantum optics, cold atomic gases, quantum dots and nuclear structure \cite{Ortiz2005}.

Neutron-proton (np) pairing has long been expected to play an important role in N=Z nuclei. both through its isovector and isoscalar character. While the relevance of isovector (J=0) np-pairing is well established, the role of isoscalar np-pairing continues is still being debated \cite{Afanasjev2013}.  The first exact solution of the proton-neutron isoscalar-isovector (T=0,1) pairing Hamiltonian with nondegenerate single-particle orbits and equal pairing strengths was presented in 2007, as a particular case of a family of integrable SO(8) Richardson-Gaudin models \cite{Lerma2007}.
There is clear evidence for an isovector np condensate as expected from isospin invariance. However, and contrary to early expectations, a condensate of deuteron-like pairs appears quite elusive \cite{Frauendorf2014}.

Particle-number-conserving formalisms have been explored for the treatment of isovector-isoscalar pairing in nuclei, but the agreement with the exact solution is less satisfactory than in the case of the SU(2) Richardson model for pairing between like particles \cite{Sandulescu2009}. This leads to the important conclusion that  the isoscalar and the isovector proton-neutron pairing correlations cannot be treated accurately by models based on a proton-neutron pair condensate \cite{Negrea2018}.

 A major step in the microscopic description of the low-energy spectrum of nuclei was made in the early 60s by Baranger \cite{Baranger1960} in terms of quasiparticle fermions with residual two-body interactions, the most important of which is the quadrupole-quadrupole interaction, associated with quadrupole deformation of the nuclear shape and the existence of rotational bands \cite{Kumar1968}. Soon thereafter this led to birth of the pairing-plus-quadruple-model \cite{Bes1969}. Mean-field quadrupole-quadrupole correlations allow for shell corrections in the single-particle structure of spherical and deformed nuclei \cite{Brack1972}. Collective properties of vibrational and rotational nuclei have been described employing boson expansion techniques with this Hamiltonian \cite{Kishimoto1976,Arima1978,Bijker1980,Isacker1981,Klein1991}.

A detailed tensorial analysis of realistic shell model Hamiltonians has conclusively shown that isovector pairing is the most important component in the particle-particle channel, while the quadrupole-quadrupole interaction is the most important in the particle-hole channel \cite{Dufour1996}. Extensive shell model calculations confirm the dominance of the quadrupole-quadrupole component of the interaction \cite{Zuker1995,Caurier2005}.

The quadrupole-quadrupole interaction can be associated with the second order Casimir operator of the SU(3) algebra, allowing for an algebraic description of nuclear dynamics \cite{Elliot1958,Hecht1973} in the rotational regime, generalized for heavy nuclei in the pseudo SU(3)model \cite{Ratna1973,Troltenier1995}, with the pseudo-spin symmetry that enters in this model now known to have a relativistic origin \cite{Blokhin1995,Meng1998,Ginocchio2005,Liang2015}. It has been found that the SU(3) symmetry of the quadrupole term is broken mainly by the one-body spin-orbit term, but that the energies depend strongly on pairing \cite{Lerma2011}. The inclusion of a quadrupole-pairing channel allowed for a very detailed description of rotational bands in heavy deformed nuclei with many quasiparticle excitations in the Projected Shell Model \cite{Hara1995,Sun2000}. While these methods were generalized to include realistic effective interactions in Hartree-Fock-Bogolyubov calculations \cite{Decharge1980}, and in extensive analysis of shape coexistence in nuclei \cite{Wood1992}, in this review we will focus on the pairing-plus-quadrupole Hamiltonian, but with the inclusion of isoscalar pairing as well for the reasons noted in the Introduction.

\subsection{Shell model description of Gamow-Teller strengths in $2p1f$-shell nuclei}

The growth in computational power, the development of sophisticated shell model codes and the use of realistic potentials, consistent with two-nucleon data, has opened in  the last decades new avenues for a detailed microscopic description of the dynamics of medium mass nuclei, both its single-particle and collective character. In particular, it enabled a quantitative description of rotational motion and Gamow-Teller transitions \cite{Caurier2005}. Full shell model calculations in the $2p1f$ shell, including the orbitals $f7/2, p3/2, p1/2$ and $f5/2$ successfully reproduce the spectra, binding energies, quadrupole moments and transition rates \cite{Poves1981,Honma2005}, providing a microscopic description of the onset of rotational motion  \cite{Caurier1995c,Caurier1995d,Zuker1995} and the quenching of  Gamow-Teller transitions \cite{Caurier1995,Caurier1995b,Nakada1996}. More recently, novel realistic interactions, like the slightly monopole-corrected version of the well-known KB3 interaction, denoted as KB3G, were able to reproduce the measured Gamow-Teller strength distributions and spectra of  $2p1f$-shell nuclei in the mass range A = 45-65  \cite{Caurier1999}, while a new shell model interaction, GXPF1J, has been employed to describe the electron capture reaction rates, and the strengths and energies of the Gamow-Teller transitions in the even isotopes of Ni from $56-64$  \cite{Suzuki2011,Cole2012}. These interactions have been successfully employed \cite{Kumar2016,Kumar2016b} to describe the most recent experimental results \cite{Fujita2005,Adachi2006,Fujita2014,Molina2015}, and have enabled a study of the evolution of the GT strength distribution from stable nuclei to very neutron-rich nuclei \cite{Yoshida2018}. In the $2p1f$ shell the estimated quenching factor is $q=0.744$, slightly smaller but statistically compatible with the $2s1d$-shell value \cite{Martinez1996}.

An anticorrelation between between the total Gamow-Teller strength and the transition rate of the collective quadrupole excitation has been observed \cite{Auerbach1993,Troltenier1996}, which can be simulated with artificial changes of the spin-orbit splitting \cite{Zelevinsky2017}.

Full shell model Monte Carlo calculations for N=Z $2p1f$-shell nuclei \cite{Radha1997} with a schematic Hamiltonian containing isovector pairing and quadrupole-quadrupole interactions  found a transition with increasing temperature from a phase of isovector pairing dominance to one where isoscalar pairing correlations dominate \cite{Langanke1997}. The appearance of T=1 ground states in N=Z odd-odd nuclei has been connected to the combined effect of the isoscalar and isovector L=0 pairing components of the effective nucleon-nucleon interaction \cite{Poves1998}.
While in general  isovector pairing dominates in the ground states, the isoscalar pair correlations depend strongly on the spin-orbit splitting \cite{Martinez1999}.  The isoscalar (T = 0, S = 1) neutron-proton pairing interaction plays a decisive role for the concentration of GT strengths at the first-excited 1+ state in $^{42}$Sc, but this effect is suppressed in heavier N=Z nuclei by the spin-orbit force supplemented by the quadrupole-quadrupole interaction \cite{Kaneko2018}. The isoscalar pairing interaction enhances the GT strength of lower energy excitations in N=Z nuclei \cite{Bai2013}. The competition between T=1 and T=0 pairing correlations was also studied  using self-consistent Hartree-Fock-Bogoliubov (HFB) plus quasiparticle random-phase approximation calculations, showing that it can cause the inversion of the $J^{\pi} = 0^+$ and $J = 1^+$ states near the ground state \cite{Sagawa2016}. An analysis of Gamow-Teller transitions and neutron-proton-pair transfer reactions reveals that the SU(4)-symmetry limit is not realized in $^{42}$Sc \cite{Isacker2018} and it is strongly broken by the spin-orbit interaction and by increasing neutron excess \cite{Petermann2007}.

\section{Systematic model calculations of Gamow Teller transitions in the $2p1f$ major shell}

In this section and in the remainder of this review we describe model calculations through which we can systematically appraise the role of the various key components of the nuclear Hamiltonian introduced in Subsection II.C on properties of Gamow-Teller transition rates (and to a lesser extent the associated nuclear spectra) for nuclei in the vicinity of N=Z. For such nuclei, we can readily focus on the interplay between all of the key interactions discussed and especially between isovector and isoscalar pairing correlations.

\subsection{Model and Optimal Hamiltonian}

 We focus our attention on even-mass nuclei in the region from A=42-48, in which the valence neutrons and protons reside outside an assumed doubly-magic $^{40}Ca$ core and are restricted to the orbitals of the $2p1f$ major shell. The Hamiltonian we use is

\begin{equation} \label{model2}
\begin{split}
\widehat{H}_{2}=\sum_{i} \varepsilon_{i} \widehat{n}_i + \chi \left(  :\widehat{Q}\cdot \widehat{Q}: + a \widehat{P}^{\dagger} \cdot \widehat{P} + b \widehat{S}^{\dagger} \cdot \widehat{S} \right).
\end{split}
\end{equation}

Here $\widehat{Q} = \widehat{Q}_{n} + \widehat{Q}_{p}$ is the quadrupole mass operator and $:\widehat{Q}\cdot \widehat{Q}:$ is the two-body part of the quadrupole-quadrupole operator. Also $\widehat{P}^{\dagger}$ is the operator that creates a correlated pair with $L=0$, $S=1$, $J=1$, $T=0$, whereas  $\widehat{S}^{\dagger}$ is the operator that creates a correlated pair with $L=0$, $S=0$, $J=0$, $T=1$. Finally, the first term is the contribution of single-particle energies, which are taken from the realistic  interaction  $KB3$ (see \cite{Poves1981}): $\varepsilon_{7/2}=0.0MeV$, $\varepsilon_{3/2}=2.0MeV$, $\varepsilon_{1/2}=4.0MeV$ and $\varepsilon_{5/2}=6.5MeV$.

Note that our Hamiltonian indeed contains the key quadrupole-quadrupole interaction emphasized in Subsection II.3 and also contains both isovector and isoscalar pairing, whose relative importance, as we noted, will be a key focus of our interest. It of course also includes an underlying  single-particle field, which as also noted in Subsect. II.3 can play a critical role.

It is interesting to note here that much the same Hamiltonian was used in Ref. \cite{Lei2011}, which also systematically explored some features of the same nuclear region. However, in that work a different set of single-particle energies was used, involving a sum of the one-body parts of the quadrupole-quadrople interaction and the spin-orbit interaction. We will briefly show the corresponding results for such a choice of Hamiltonian in Appendix A.

We will focus here on the effect of varying the strength parameters for the isovector and isoscalar pairing terms, leaving the strength of the quadrupole-quadrupole interaction and the single-particle energies unchanged. As noted above, we restrict the analysis to even-mass nuclei near the beginning of the $2p1f$ shell, namely with $A=42-48$. All calculations reported here have been carried out using the ANTOINE shell-model code \cite{Antoine,Caurier1999b,Caurier2005}.

 As our goal is to leave the quadrupole-quadrupole strength and the single-particle energies unchanged as we search for the effect of varying the two pairing strengths, we need to first choose an optimal set of parameters. We can then vary the pairing strength parameters away from their optimal values, while keeping the others fixed,  to see how these changes impact the description of the properties of interest. Our approach for choosing the {\it optimal} Hamiltonian parameters is based on two primary criteria:

\begin{enumerate}
\item Good reproduction of the low-energy spectra of the nuclei of interest (especially the $1^+$ states of odd-odd nuclei),
\item Good description of GT properties and especially their fragmentation.
\end{enumerate}

Our analysis, described in some detail in Appendix B, suggests:
\begin{itemize}
\item
For $N=Z$ nuclei the optimal set of parameters for even-mass nuclei with $A=42-48$ are:  $ \chi=-0.065\ MeV$, $a=b=6 $.
\item
 When $ N \neq Z $, as likewise discussed in Appendix B,  the ground states of odd-odd nuclei with such a Hamiltonian and attractive isoscalar pairing typically have spin and parity $1^+$, at variance with the experimental data. Thus, for $N \neq Z$ nuclei, we simply {\it turn off} the isoscalar pairing, {\it i.e.} set $a=0$. This leads to a good overall description of most features exhibited by the nuclei we consider, thereby providing the starting part for our analysis of the impact of changes to the pairing strengths.
 \end{itemize}

In what follows we first very briefly discuss the energy spectra emerging from this optimal Hamiltonian and then focus on the role of the various pairing modes on GT transitions. A further discussion of the role of the pairing modes on energy spectra is reserved for Appendix B.

\section{Results}

\subsection{Energy spectra}

\begin{figure}[hbtp]
\centering
\includegraphics[width=1.0\textwidth]{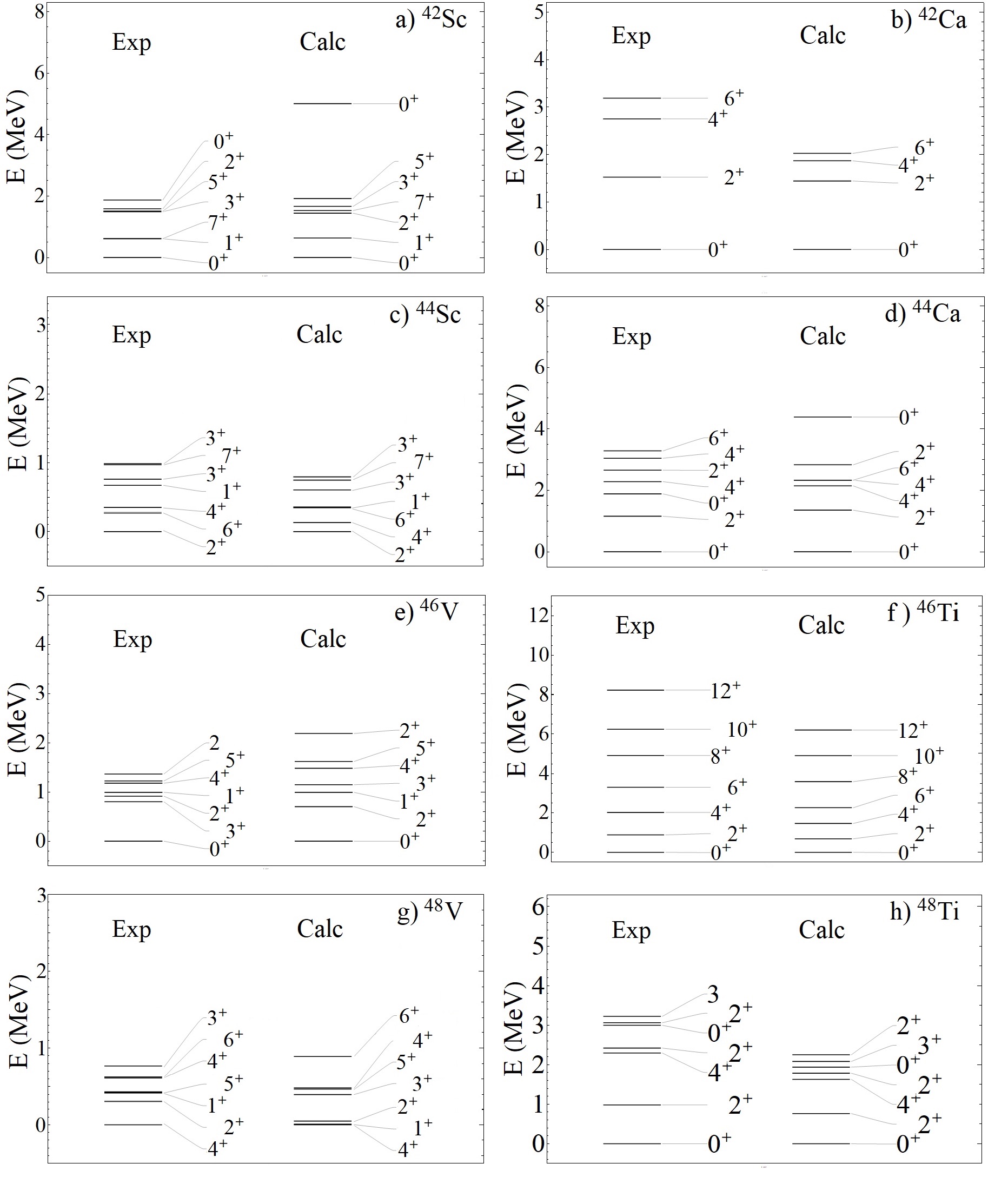}
\caption{Energy spectra of the daughters (a, c, e and g) and parents (b, d, f and h) nuclei, comparing the  results obtained with the optimal Hamiltonian to experimental data \cite{exp}.}
\end{figure}

The energy spectra that emerge from our optimal Hamiltonian for the eight nuclei involved in the GT transitions we study are shown in Fig. 1.
All experimental data was obtained from \cite{exp}.
In all cases the ground state angular momentum agrees with the observed values. Furthermore, in all odd-odd daughter nuclei the energies of the first $ 1^{+}$ are accurately reproduced by the model. Lastly,  the energy of the first $ 2^{+}$ state in all of the nuclei we consider is fairly well reproduced.

In the case of $ ^{48}V$ the model predicts a near degeneracy between the lowest three states, those with $J^{\pi}$=$0^+$, $1^+$ and $2^+$, somewhat closer than in experiment, but nonetheless in reasonable agreement.

On the other hand, the calculated energies of states in even-even nuclei with angular momenta higher than $2^+$ are somewhat more compressed than the experimental energies, thereby losing the rotational properties that were better described in Ref. \cite{Lei2011}. In Appendix B, we show how increasing the pairing strengths expands these energy spectra, however at the cost of losing the fragmentation of GT intensities that emerges nicely for the optimal Hamiltonian (see the following subsection).

\subsection{GT Transitions}

Below are the results obtained for GT transition intensities, and their dependence on the pairing strengths. The values of B(GT) are multiplied by the usual quenching factor $(0.74)^{2}$ \cite{Martinez1996,Gysbers2019}.

\subsubsection{A=42}

The GT transition strengths for  $^{42}Ca$ $\rightarrow$ $^{42}Sc$ are shown in Fig. 2, as functions of the isovector pairing strength $b$, which applies to both the parent and daughter nucleus, {\it and} the isoscalar pairing strength $a$, which only contributes to the properties of the $^{42}Sc$ daughter nucleus.

\begin{figure}[hbtp]
\centering
\includegraphics[width=1.0\textwidth]{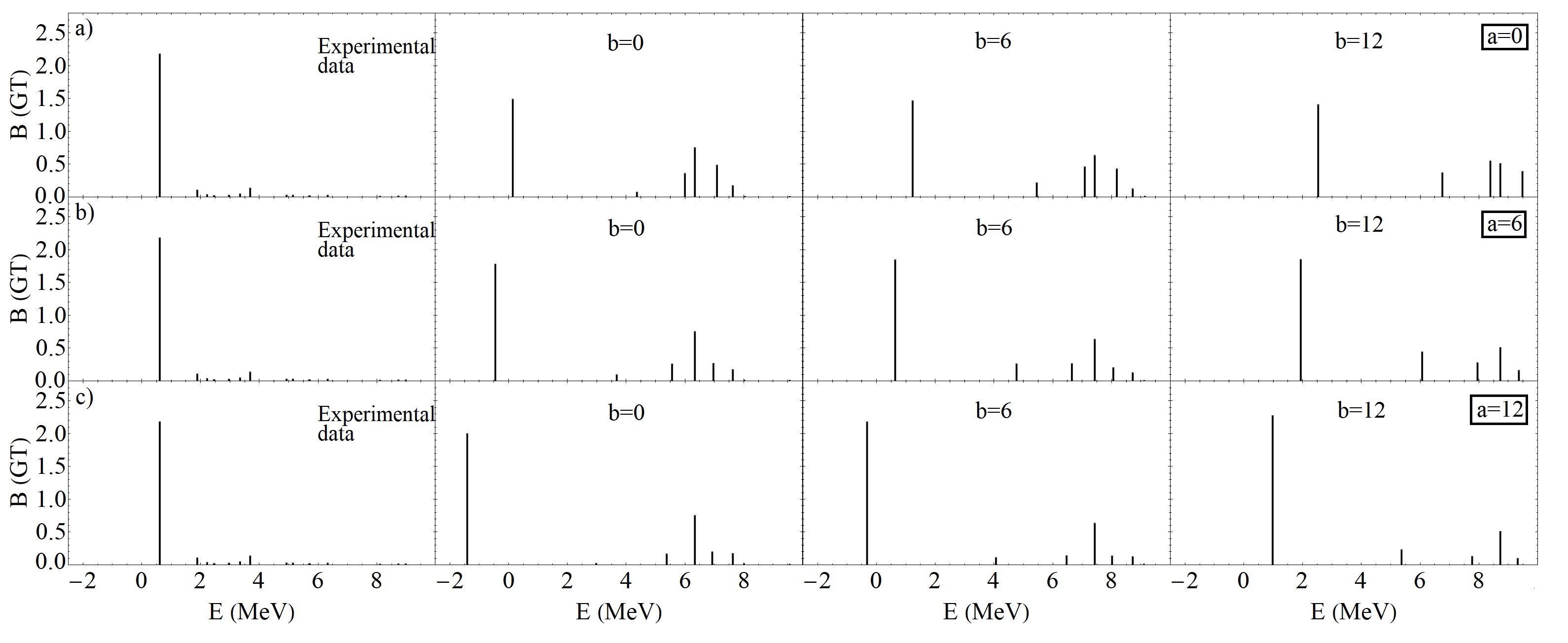}
\caption{Comparison of the experimental \cite{Fujita2015} and theoretical results for  B(GT) transition strengths for $^{42}Ca$ $\rightarrow$ $^{42}Sc$ as a function of the isovector pairing strength $b$ and the isoscalar pairing strength $a$, which only acts in the daughter nucleus. }
\end{figure}

When $a=b=6$ (the optimal pairing strengths) there is a single strong peak at almost the exact experimental energy and strength.  There are small satellite peaks at higher energy with somewhat more strength than seen experimentally.

As $a$ is increased the strength to the lowest $1^+$ state increases, albeit slowly, while the energy of that state goes down in energy and eventually for $a=12$ becomes the ground state. As $b$ is increased, for a given $a$, the main peak moves up in energy, but with no noticeable change in its strength. The presence of a single dominant peak is an indication that these nuclei have good SU(4) symmetry.

\subsubsection{A=44}

Fig. 3 depicts the GT intensities for $^{44}Ca$ $\rightarrow$ $^{44}Sc$. As both nuclei have a neutron excess, we set $a=0$ for both and vary the isovector pairing strength $b$ only.

\begin{figure}[hbtp]
\centering
\includegraphics[width=1.0\textwidth]{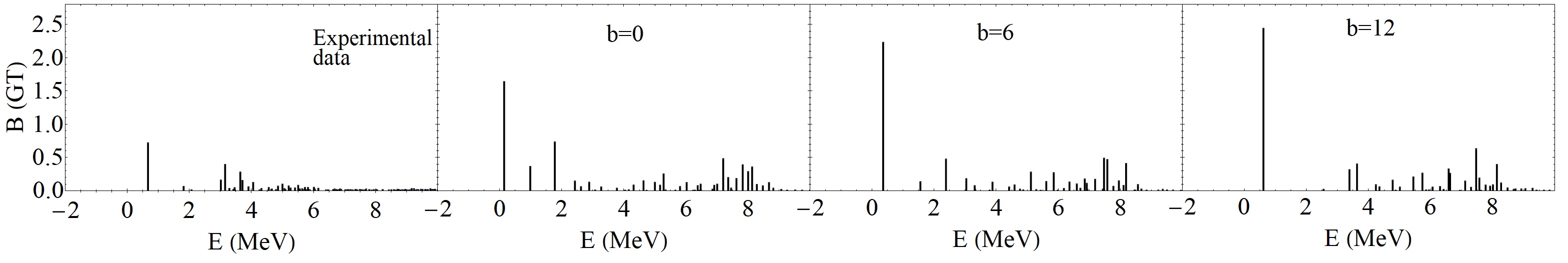}
\caption{Comparison of the experimental \cite{Fujita2013} and theoretical results for B(GT) transition strengths for $^{44}Ca$ $\rightarrow$ $^{44}Sc$ as a function of the isovector pairing strength $b$. Since both nuclei involved have neutron excesses, the isoscalar pairing strength is set to $a=0$.  }
\end{figure}

In this case, and in contrast to $A=42$, the experimental data shows several satellite peaks at fairly low energies. For the optimal $b=6$ isovector strength, the lowest excitation is roughly five times more strongly populated than the next few, in contrast with the experimental data where the relative enhancement is roughly two. The lowest peak moves up in energy as $b$ is increased and becomes progressively more dominant.

Though not part of the A=44 GT decay, it is worth commenting briefly here on the N=Z nucleus $^{44}Ti$. The spectrum of $^{44}Ti$ is shown in Figure 11 of Appendix B.2 for several choices of the isovector and isocalar pairing strengths (since it has N=Z both pairing modes are relevant). The first point to note is that like $^{44}Sc$ and $^{44}Ca$ the lowest states are well described by the optimal Hamiltonian. In contrast to them, however,  the higher angular momentum states are too compressed, so that the resulting spectrum is even more removed from that of an SU(3) rotor. We can partially restore the SU(3) pattern and get a better overall description description of the energy spectrum by increasing the isovector pairing strength to $b=12$. On the other hand, as we see from Fig. 3, an increase in the isovector pairing strength would lead to a loss of fragmentation in the corresponding mass-44 GT pattern, in worse agreement with the experimental data.  Thus there is a competition between the fragmentation produced by the spin-orbit operator \cite{Petermann2007} and by isovector pairing.

\subsubsection{A=46}

Next, we analyze the GT results obtained for the nuclei with mass $ A = 46 $. While the parent nucleus involved in the GT decay $^{46}Ti$ $\rightarrow$ $^{46}V$ has a neutron excess, the daughter nucleus does not. Thus we assume $a=0$ for the parent nucleus and present the results as a function of the $a$ value used in describing the daughter. In addition, the results are shown as a function of the isovector pairing strength $b$ used for both nuclei. These results are shown in Fig. 4.

\begin{figure}[hbtp]
\centering
\includegraphics[width=1.0\textwidth]{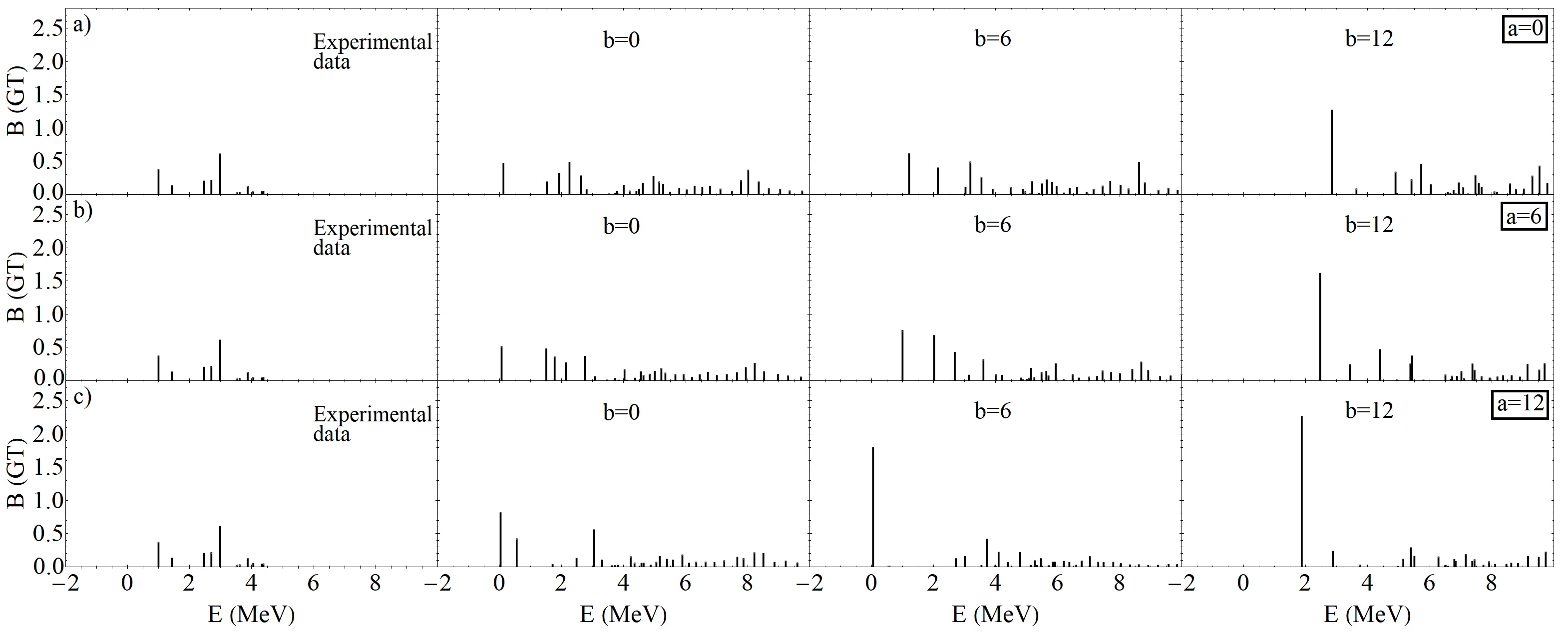}
\caption{Comparison of the experimental \cite{Adachi2006} and theoretical results for B(GT) transition strengths for $^{46}Ti$ $\rightarrow$ $^{46}V$ as a function of the isovector pairing strength $b$ and the isoscalar pairing strength $a$, which only acts in the daughter nucleus. }
\end{figure}

The results are interesting. Here there is strong fragmentation of the strength for the optimal $a=b=6$ parameters, though the lowest state has slightly more strength than the next few. However, the overall strength to these states is in reasonable accord with what is seen in experiment.

Other features of the results worth noting are that (1) even though we produce an appropriate fragmentation pattern the individual strengths for $a=b=6$ are substantially larger than in the data, (2) the effect of increasing $a$ is, as for the other masses studied, to focus increasing strength in the lowest $1^+$ state while lowering its energy and (3) the effect of increasing $b$ is, again as for other masses, to likewise enhance population of the lowest $1^+$ state but now while lifting its energy.

As in our mass-44 discussion, we also comment here on the rotational energy pattern for the even-even nucleus $ ^{46}Ti$. As can be seen from  Fig. 1 f, the higher angular momentum states are compressed with respect to experiment for the optimal Hamiltonian and would require an increase in isovector pairing (see Fig. 13 in Appendix B.3) to improve the overall description of those states and in doing so partially restore the rotational SU(3) symmetry.  But as just noted,  such an increase of the isovector pairing strength has the effect of focussing the GT strength in the lowest peak, thereby also serving to partially restore SU(4) symmetry.

Though isoscalar pairing is only of importance to the daughter nucleus $^{46}V$, its increase likewise has the effect of restoring SU(4) symmetry, as is evident from Fig. 4, and to worsen the reproduction of the experimental results.

\subsubsection{A=48}

Finally, we treat the GT transitions in $ A = 48 $. In this case the relevant decay is $^{48}Ti$ $\rightarrow$ $^{48}V$, for which both the parent and daughter nuclei have a neutron excess. Thus, in Fig. 5, where we compare the experimental and calculated transition rates, the theoretical analysis is only shown as a function of the isovector pairing strength $b$.

\begin{figure}[hbtp]
\centering
\includegraphics[width=1.0\textwidth]{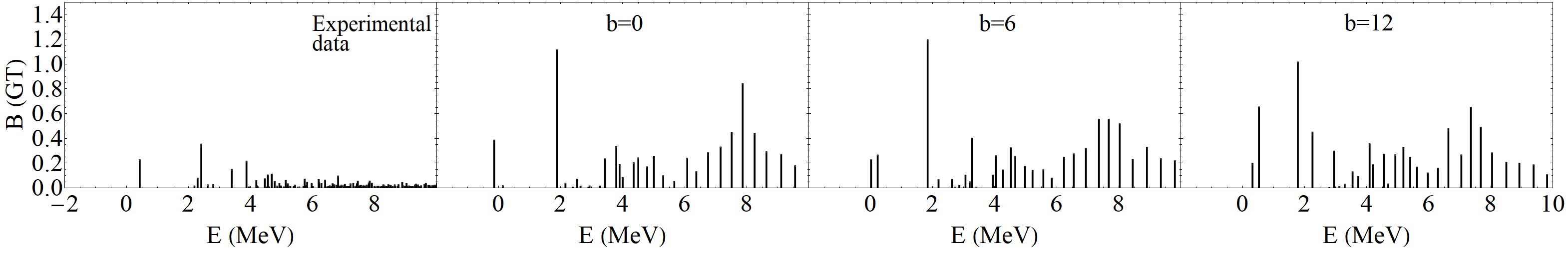}
\caption{Comparison of the experimental \cite{Ganio2016} and theoretical results for  B(GT) transition strengths for $^{48}Ti$ $\rightarrow$ $^{48}V$ as a function of the isovector pairing strength $b$. Since both nuclei involved have neutron excesses, the isoscalar pairing strength is set to $a=0$ for both.  }
\end{figure}

The model with the optimal $b=6$ isovector strength produces the lowest energy $1^+$ excitation in close agreement with experiment. However, the level of fragmentation in the data is not well reproduced. The strength is concentrated in a single peak at $2~MeV$, which is where the strongest state lies in the data, but it is several times more strongly populated than any other.  As $b$ is increased, the overall effect is to increase the level of fragmentation across an increasingly wider range of states, in worse the agreement with experiment.

In this case, the optimal parameters give acceptable results for the energy spectra of the daughter and parent nuclei (Fig 1g and 1h), neither of which extends to particularly high angular angular momentum. In the case of $^{48}Cr$, where the data extends to high angular momentum and whose results are shown in Figure 16 in Appendix B.4, the optimal Hamiltonian gives a good description of the spectrum up to $J^{\pi}=16^{+}$. The fact that the rotational pattern of this nucleus is fairly well described without having to dramatically modify the pairing strengths reflects the fact that in a system with a large product of the number of valence neutrons and protons $N_p \times N_n$, the effects of the quadrupole-quadrupole interaction are increasingly more dominant and changes in the pairing strengths have a less important effect.

\section{Summary and Conclusions}

In this work, we first reviewed the role played by Gamow-Teller transitions in stellar nucleosynthesis and in double beta decay, and summarized theoretical results obtained employing shell model and mean field techniques over the last half century. The pairing plus quadrupole Hamiltonian was introduced and the role of the isoscalar pairing interaction was discussed.

In the second part of this review we explored the effects of proton-neutron pairing on even-mass nuclei in the beginning of $2p1f$ shell. The analysis was done in the framework of the nuclear shell model using a parametrized Hamiltonian that contains not only isoscalar and isovector pairing but also a quadrupole-quadrupole interaction that produces background deformation.

The first step in this analysis was the choice of the optimal Hamiltonian parameters within our model. This was done by focussing on the parameters that provided an optimal description of the energies of the lowest $1^+$ states in the nuclei of interest, a good description of the other low-lying states of these nuclei, and an optimal GT fragmentation pattern.  A significant outcome of this part of the analysis was the realization that for a restricted model Hamiltonian of the type we used we must {\it turn off} isoscalar pairing when dealing with nuclei having a neutron excess to avoid producing the incorrect ground state spin and parity for many such nuclei.  Our optimal Hamiltonian is able to achieve good overall fits to experimental data for both energy spectra and GT decay properties, albeit with the limitations inherent in the relatively simple parametrization we use.  Highlights are an accurate reproduction of the properties of the lowest $1^{+} $ states,  and a reasonable description of the GT fragmentation pattern.

We then varied the isoscalar and isovector pairing strengths from their optimal values to systematically study how the two pairing modes affect first the GT properties of these nuclei and subsequently, in the Appendix, the energy systematics.  Our analysis extends from A=42 to A=48.

Our analysis of the effect on GT transition properties showed that an increase in the isoscalar pairing strength in those systems in which it is active ($N =Z$ nuclei) focuses GT strength on the lowest $1^+$ state while lowering its energy. Increasing the isovector pairing strength, which is always active, also focuses GT strength on the lowest $1^+$ state but raises its energy.

Our analysis of the impact of the two pairing modes on energy spectra showed that the isoscalar and isovector pairing modes focus primarily on the odd-J states and the even-J states, respectively. Enhancing the isoscalar strength systematically lowers the first $1^+$ state and expands the set of odd-J states. In contrast, enhancing the isovector pairing strength expands the even-J part of the spectrum.

\section*{Acknowledgements}
We acknowledge helpful advice of Alfredo Poves on the use of the code ANTOINE and Lei Yang for sharing valuable information.
This work received partial economic support from DGAPA- UNAM projects IN109417 and IN104020.

\appendix
\section{The original model}\label{A}

As mentioned earlier, the work describe in Sect. 3 took inspiration from the model introduced in \cite{Lei2011}, for which the Hamiltonian is
\begin{equation} \label{model1}
\begin{split}
\widehat{H}_{1}=\chi \left(  \widehat{Q}\cdot \widehat{Q} + a \widehat{P}^{\dagger} \cdot \widehat{P} + b \widehat{S}^{\dagger} \cdot \widehat{S} +\alpha \sum_{i} \widehat{\overrightarrow{l}_{i}}\cdot \widehat{\overrightarrow{s}_{i}} \right).
\end{split}
\end{equation}

For the sake of comparison, we show here the energy spectra that emerged for that Hamiltonian. The energy spectra of the daughters (a, c, e and g) and parents (b, d, f and h) nuclei, compared with the experimental data, are shown in Fig. 6. They are calculated with the same optimal parameters as for the model discussed in Sect, 3, namely $ \chi = -0.065 \ MeV $, $ a = b = 6 $, but with $\alpha = 20$ as was used in \cite{Lei2011}. Here too in the case of $ N \neq Z $ we must {\it turn off} the isoscalar pairing, (\i.e. set $a=0$), to produce the correct energy of the lowest $1^+$ states in those nuclei.

\begin{figure}[hbtp]
\centering
\includegraphics[width=1.0\textwidth]{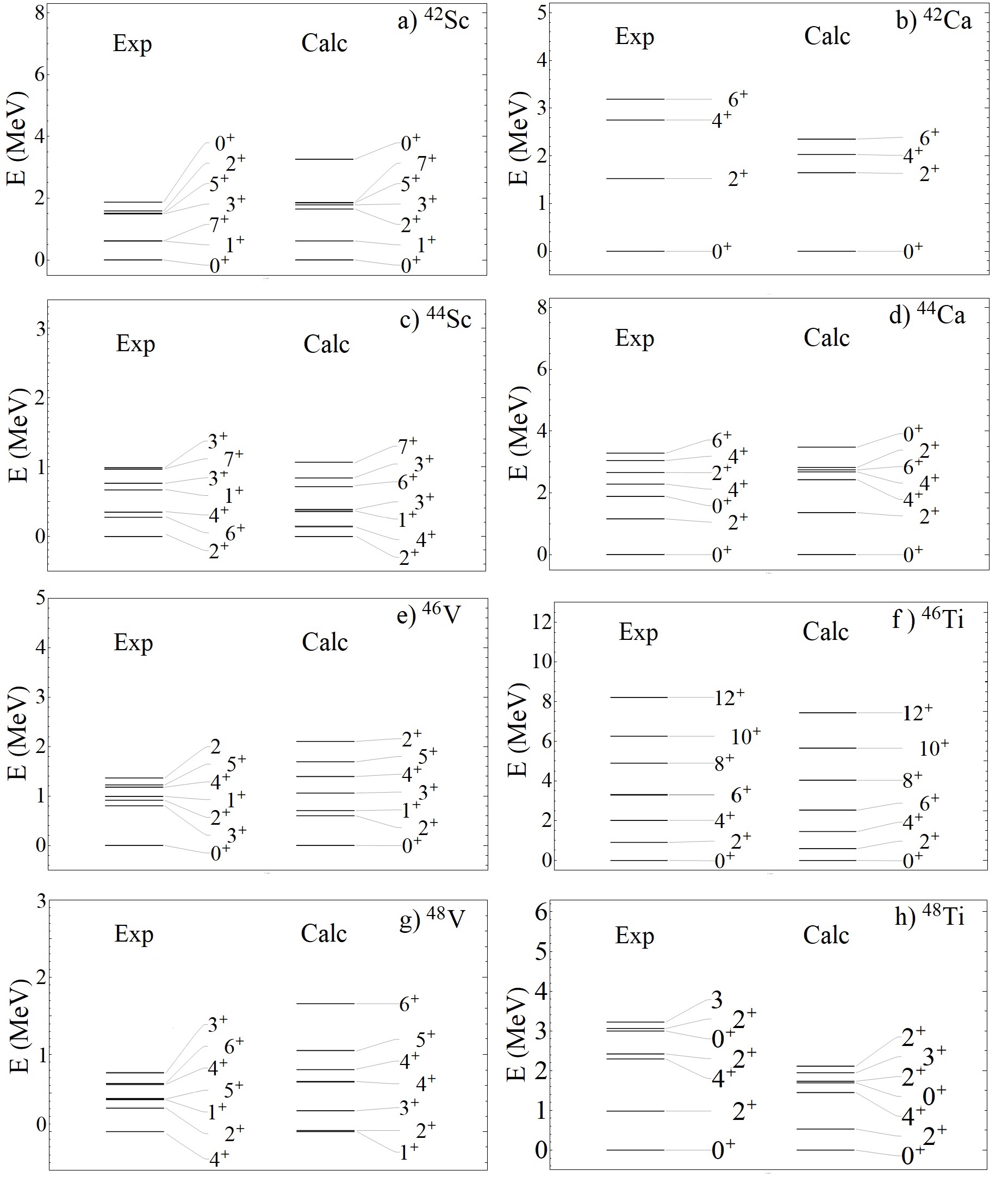}
\caption{Energy spectra of the daughters (a, c, e and g) and parents (b, d, f and h) nuclei, comparing experimental data and results obtained with the Hamiltonian (\ref{model1}).}
\end{figure}

Having the full quadrupole-quadrupole interaction, and not just its two-body part as in Eq. (1),  provides a better description of the rotational properties of these nuclei and in particular the spreading of the higher part of their spectra. On the other hand, no combination of parameters reproduces  the $4^+$ ground state in $^{48}V$. This is the main reason we have chosen to use the single-particle energies of the KB3 interaction in the results presented in Sect. 3 of the main body of this paper.

\section{Energy Spectra and rotational properties}\label{aped.B}

In this part of the Appendix, we return to the Hamiltonian described in Sect. 3 of the main body of this review and show the effect of varying the strengths of the isoscalar (where appropriate) and isovector terms in the Hamiltonian (1) on the energy spectra of the nuclei under discussion. Some of this material has been referred to already in Sect. 3 and in such cases we do not repeat the discussion.

\subsection{A=42}

In Fig. 7, results for the energy spectra for the daughter nucleus $^{42}Sc$ are shown.
\begin{figure}[hbtp]
\centering
\includegraphics[width=1.0\textwidth]{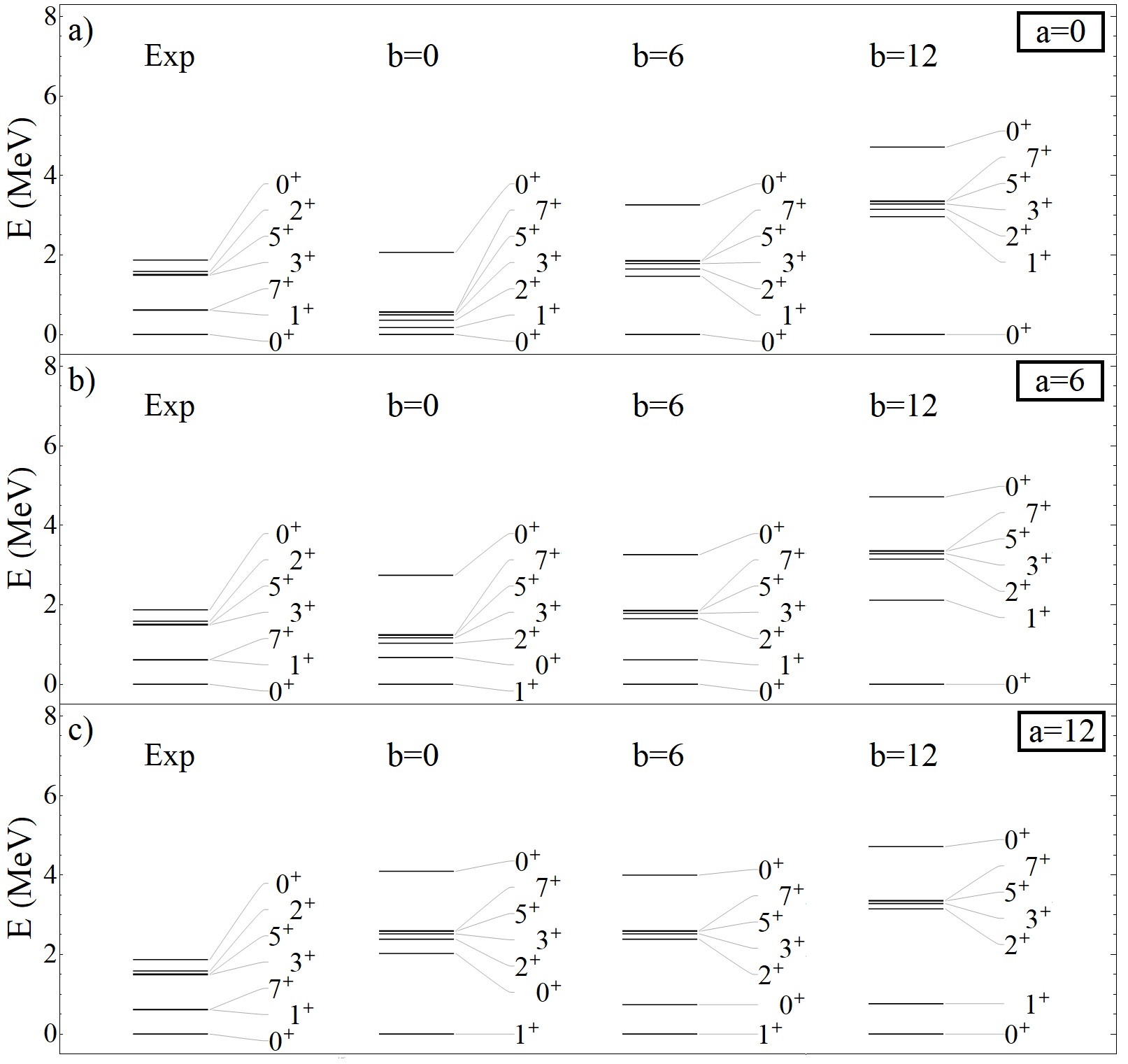}
\caption{Energy spectra of $^{42}Sc$ as a function of the isovector $b$ and isoscalar $a$  pairing strengths in comparison with the experimental data. Results are shown for $a,~b~=0,~6$, and $12$. }
\end{figure}

Two points should be noted:
(1) The best overall reproduction of the experimental spectrum occurs for equal values of $a$ and $b$, and \\
(2) the optimal choice is $a=b=6$.
With these values the lowest $1^{+}$ state has essentially the correct energy and there is a reasonable density of low-lying states. The results, though optimal for this Hamiltonian parametrization, still show clear limitations of the model. In particular, it is not possible to produce a very low-lying $7^{+}$ state, as is present in the data. It would be necessary to include other components of the two-body interaction to reproduce this.

Another point worthy of note in Fig. 7 concerns the results in the presence of pure isovector pairing. When the isovector pairing strength $b$ is weak the ground state is a $1^+$. As the isovector strength is ramped up the $0^+$ state emerges as the ground state.  Thus in the presence of non-degenerate single-particle levels it is the interplay of isovector and isoscalar pairing that produces the ground state spin in $N=Z$ nuclei, a conclusion that in fact carries through to heavier $N=Z$ nuclei as we will see when we present those results.

Next we consider the corresponding results for $^{42}Ca$, the parent nucleus for GT decay. These results are shown in Fig. 8. We only present the levels of the ground state rotational band in these figures.

\begin{figure}[hbtp]
\centering
\includegraphics[width=1.0\textwidth]{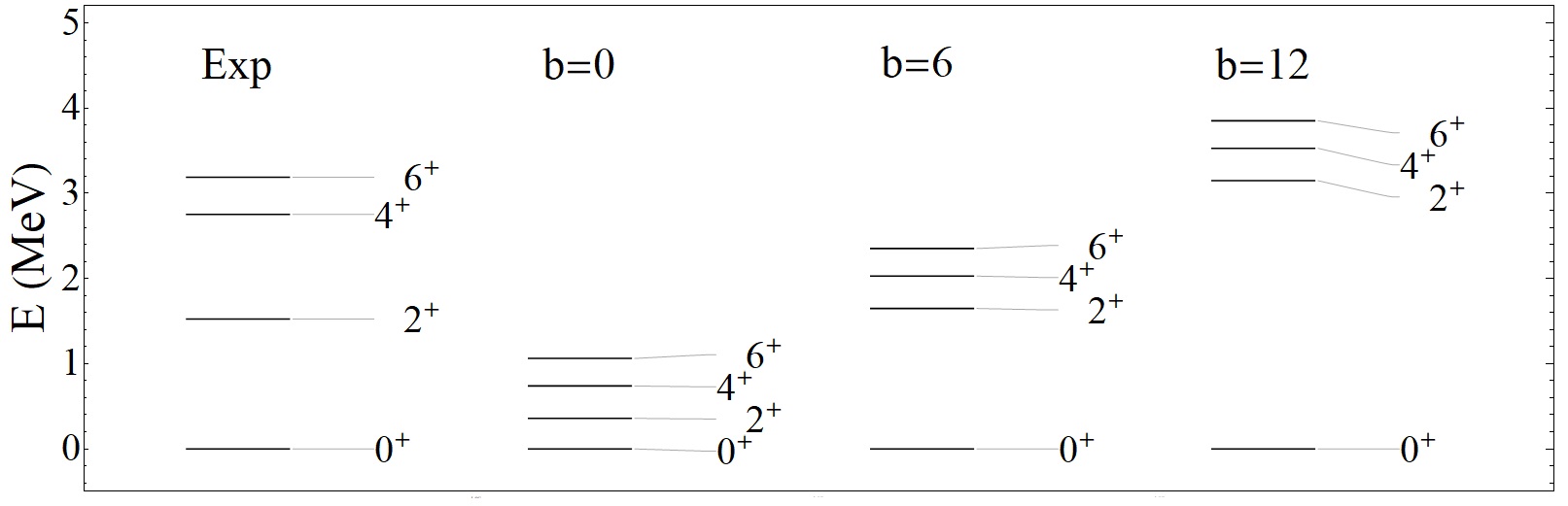}
\caption{Rotational band of $^{42}Ca$ as a function of the strength of the isovector pairing interaction $b$ in comparison with the experimental data.}
\end{figure}

Since there are only valence neutrons in $^{42}Ca$ isoscalar pairing is omitted from the figure. Note that the spectrum does not exhibit a rotational pattern even for $b=0$. Clearly the presence of the single-particle energies erases the SU(3) rotational behavior of the purely quadrupole interaction, and in the presence of isovector pairing the rotational pattern is not recovered. In all cases, the $2^+$, $4^+$ and $6^+$ tend to group together, in marked contrast to the experimental data where only the $4^+$ and $6^+$ states are grouped.
Nevertheless the optimal results seem to arise when we choose $b=6$, in accord with the conclusion from the $^{42}Sc$ analysis.

\subsection{A=44}

The calculated spectra of the daughter $^{44}Sc$ are shown in Fig. 9 with the experimental data..

\begin{figure}[hbtp]
\centering
\includegraphics[width=1.0\textwidth]{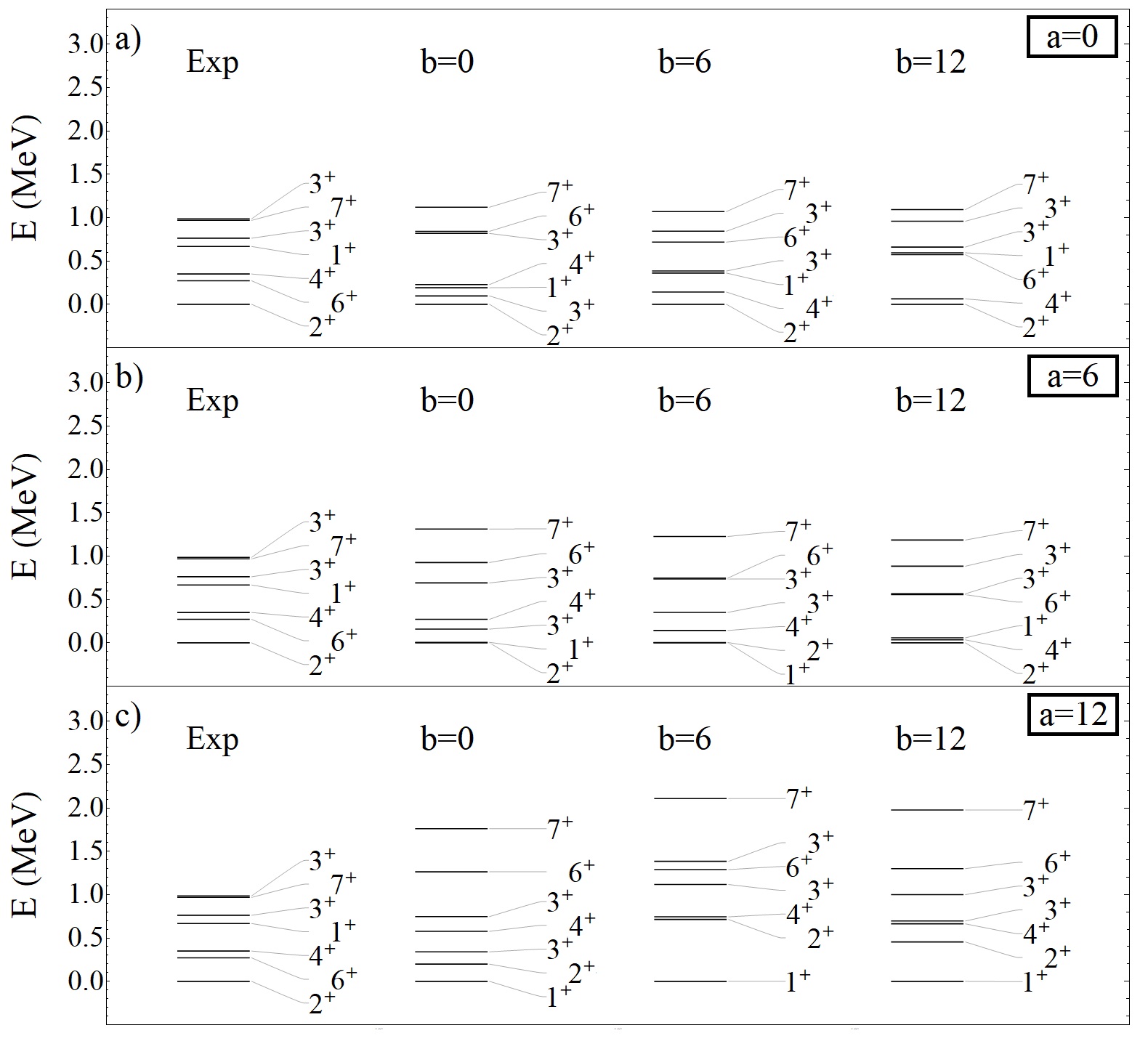}
\caption{Energy spectra of $^{44}Sc$ as a function of the isovector $b$ and isoscalar $a$  pairing strengths in comparison with the experimental data. Results are shown for $a,~b~=0,~6$, and $12$.}
\end{figure}

We include isoscalar pairing in the figure as it enables us to {\it demonstrate}  that isoscalar pairing must be removed when $N \neq Z$.
Notice that only in the case $a=0$ is the ground state spin  $2^+$ reproduced. Using $a=6$ or $a=12$ the ground state spin and parity are $1 ^{+} $.

The results for the parent nucleus $^{44}Ca$ are shown in Fig. 10. The features observed as a function of the isovector strength parameter are similar to those seen in $^{42}Ca$: the optimal energy spread arises for $b=6$, the spectrum is compressed as the isovector strength is decreased with the gap between the ground state and the first excited $2^+$ state becoming very small, and the first excited $0^+$ state remains too high in energy for all isovector strengths.

\begin{figure}[hbtp]
\centering
\includegraphics[width=1.0\textwidth]{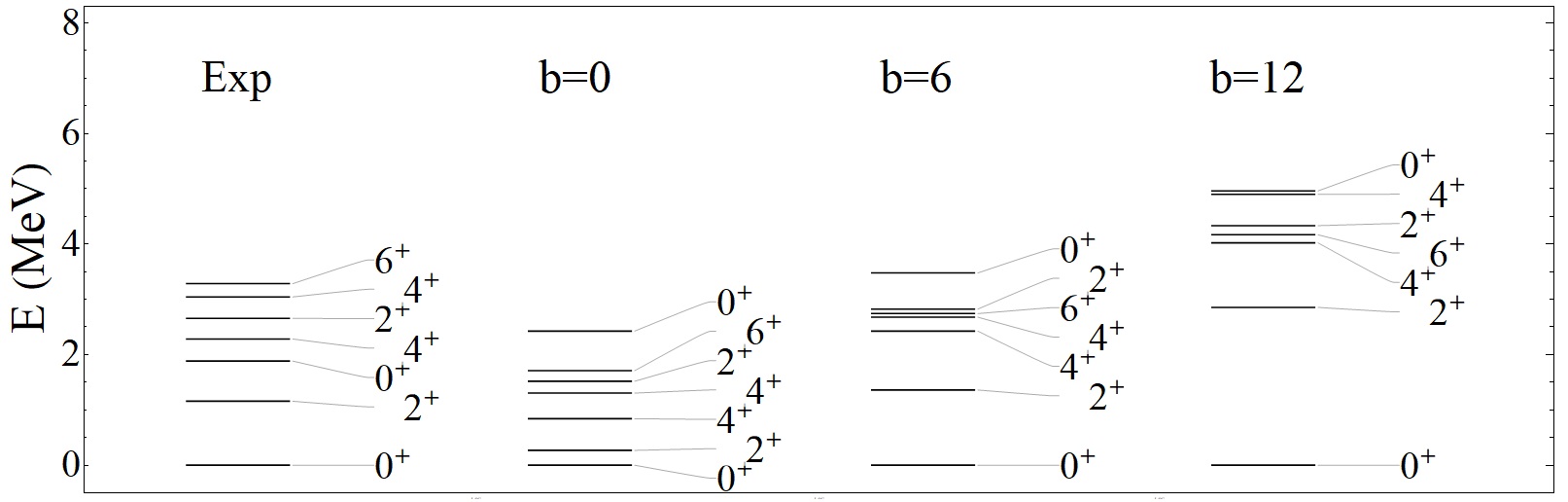}
\caption{ Energy spectra of $^{44}Ca$ as a function of the isovector pairing strength b in comparison with the experimental data.}
\end{figure}

\begin{figure}[hbtp]
\centering
\includegraphics[width=1.0\textwidth]{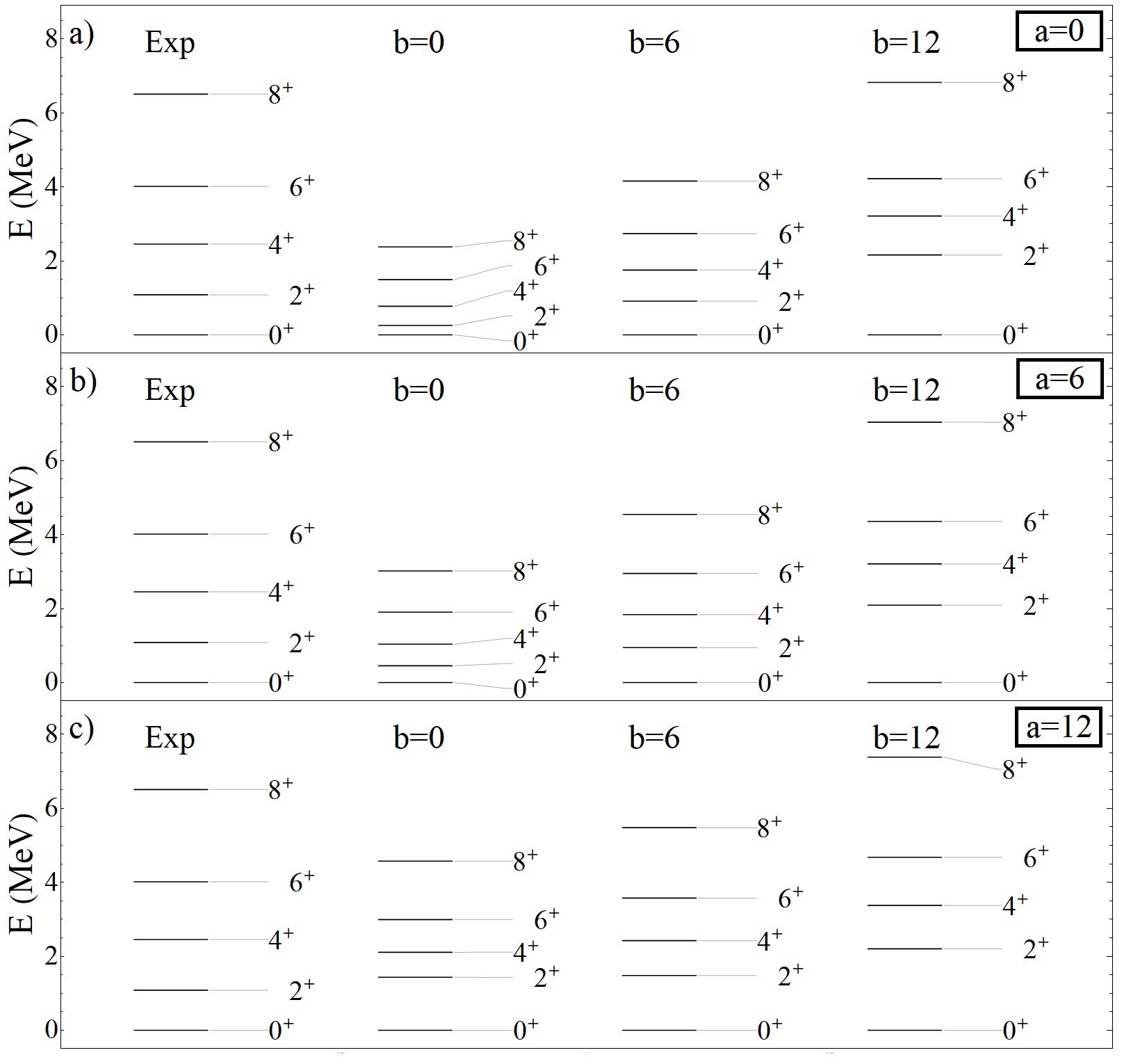}
\caption{Rotational band of $^{44}Ti$ as a function of the isovector $b$ and isoscalar $a$  pairing strengths in comparison with the experimental data. Results are shown for $a,~b~=0,~6$, and $12$.}
\end{figure}

Next we consider $^{44}Ti$ with 2 valence neutrons and 2 valence protons, for which the results are shown in Fig. 11.  Though not a part of the GT decay, it is worth looking at it nevertheless to study the effect of the different pairing modes on its rotational properties. Without pairing and single-particle effects the quadrupole-quadrupole interaction would produce a pure rotor with SU(3) symmetry. Single-particle effects by themselves perturb this pattern, as can be seen from the $a=b=0$ results in the figure. Both isovector and isocalar pairing spread the spectrum, in better agreement with experiment, but do not restore the pure rotational character.

\subsection{A=46}

Now we consider the energy spectra of the $A=46$ nuclei involved in the GT decay $^{46}Ti$~$\rightarrow$ ~$^{46}V$.

The results for $^{46}V$  are shown in Fig. 12.
The same basic features as for lighter $N=Z$ nuclei are seen here. When the intensity of both pairings decrease simultaneously, the spectrum energies are reduced.
When we reduce the isoscalar pairing alone (note this is an $N=Z$ nucleus so that isoscalar pairing is relevant), there is no appreciable effect on the states with even angular momentum, whereas those with  odd angular momentum gradually compress. Finally, when decreasing the isovector strength alone, a compression of the overall spectrum occurs and the ground state gradually changes from $ 0^{+} $ to $ 1^{+} $.

\begin{figure}[hbtp]
\centering
\includegraphics[width=1.0\textwidth]{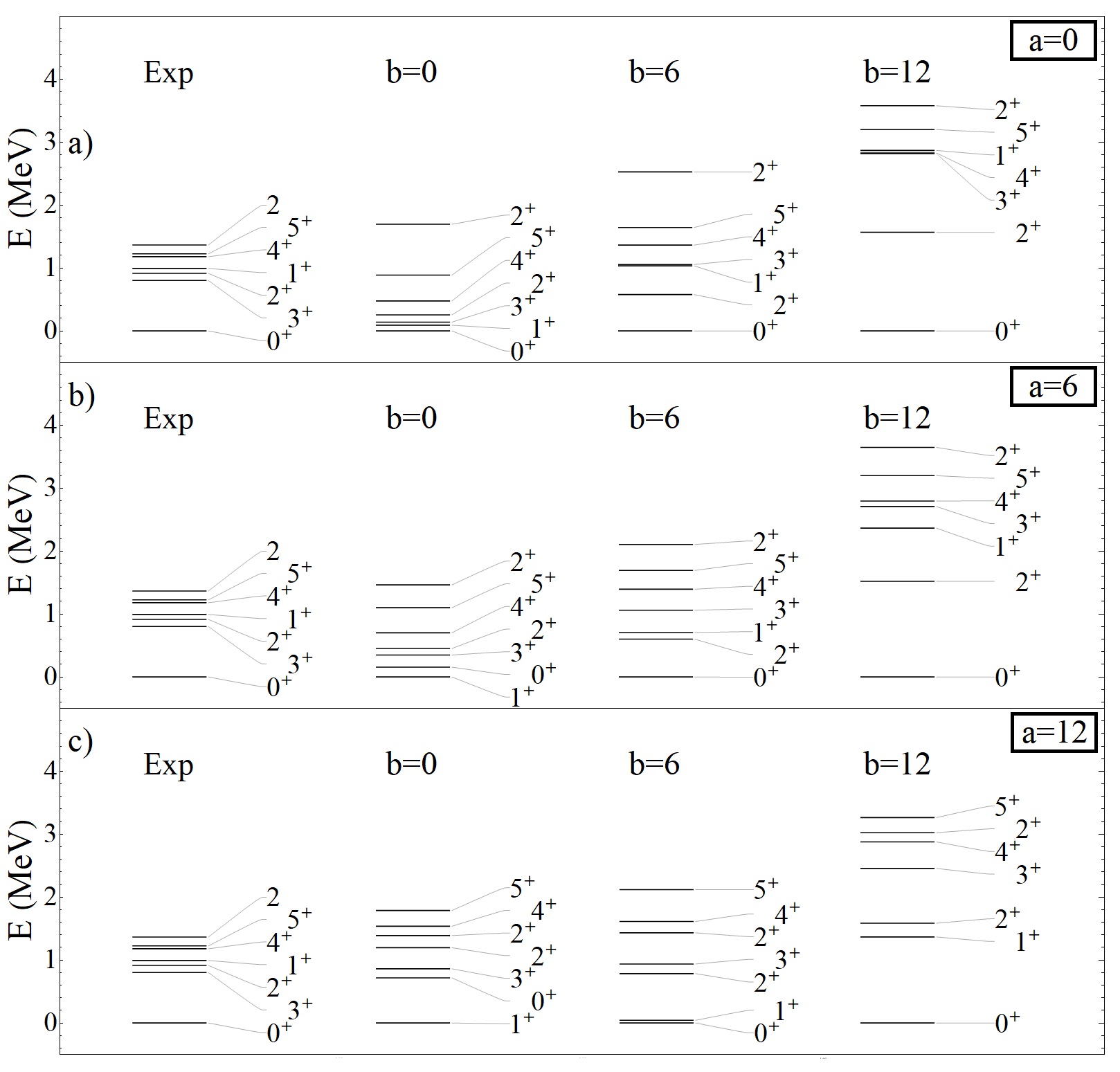}
\caption{Energy spectra of $^{46}V$ as a function of the isovector $b$ and isoscalar $a$  pairing strengths in comparison with the experimental data. Results are shown for $a,~b~=0,~6$, and $12$.}
\end{figure}

Next we turn our discussion to $^{46}Ti$ for which the relevant results for the rotational band are shown in Fig. 13. As a nucleus with a neutron excess we only show the results as a function of $b$. The main features of these results were already described in Section 3.2.3.
\begin{figure}[hbtp]
\centering
\includegraphics[width=1.0\textwidth]{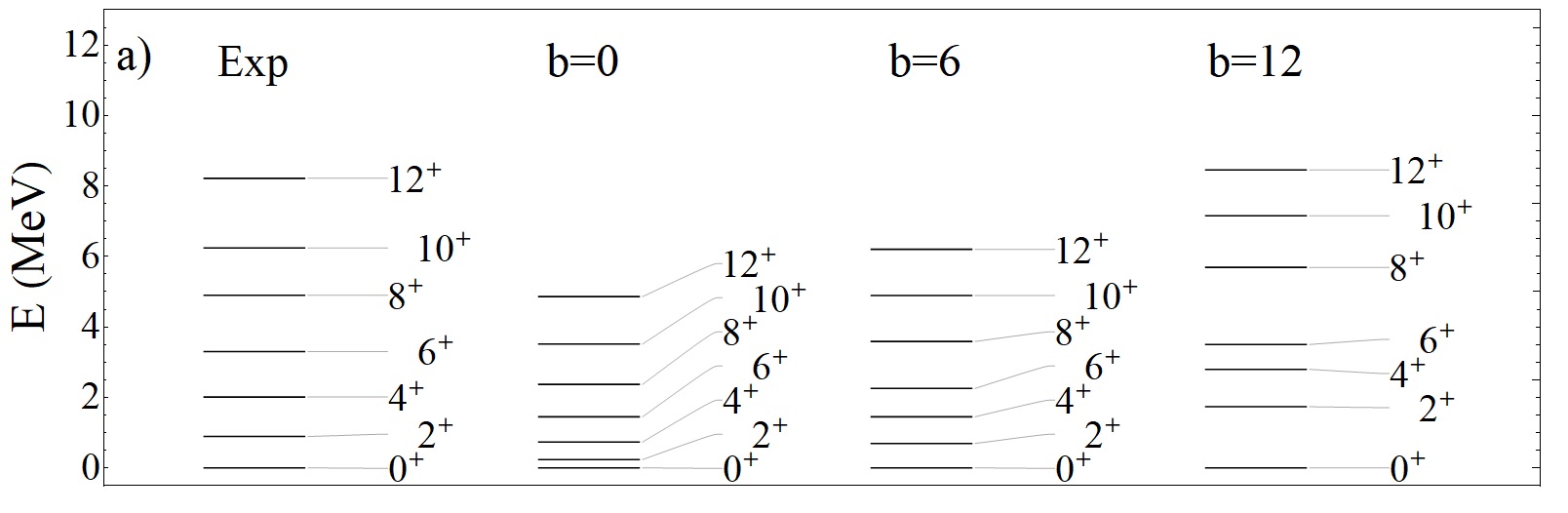}
\caption{Rotational band of $^{46}Ti$ as a function of the isovector $b$ pairing strength in comparison with the experimental data. Results are shown for $a = 0,~b~=0, 6$, and $12$.}
\end{figure}

\subsection{A=48}

Lastly we treat nuclei with $A=48$. We first discuss those that participate in the GT decay $^{48}Ti$ $\rightarrow$ $^{48}V$ and afterwards briefly comment on $^{48}Cr$.

The results for $^{48}V$ are shown in Fig. 14. We include $a \neq 0$ values even though this nucleus has a relatively large neutron excess, for reasons that will be made clearer soon.

\begin{figure}[hbtp]
\centering
\includegraphics[width=1.0\textwidth]{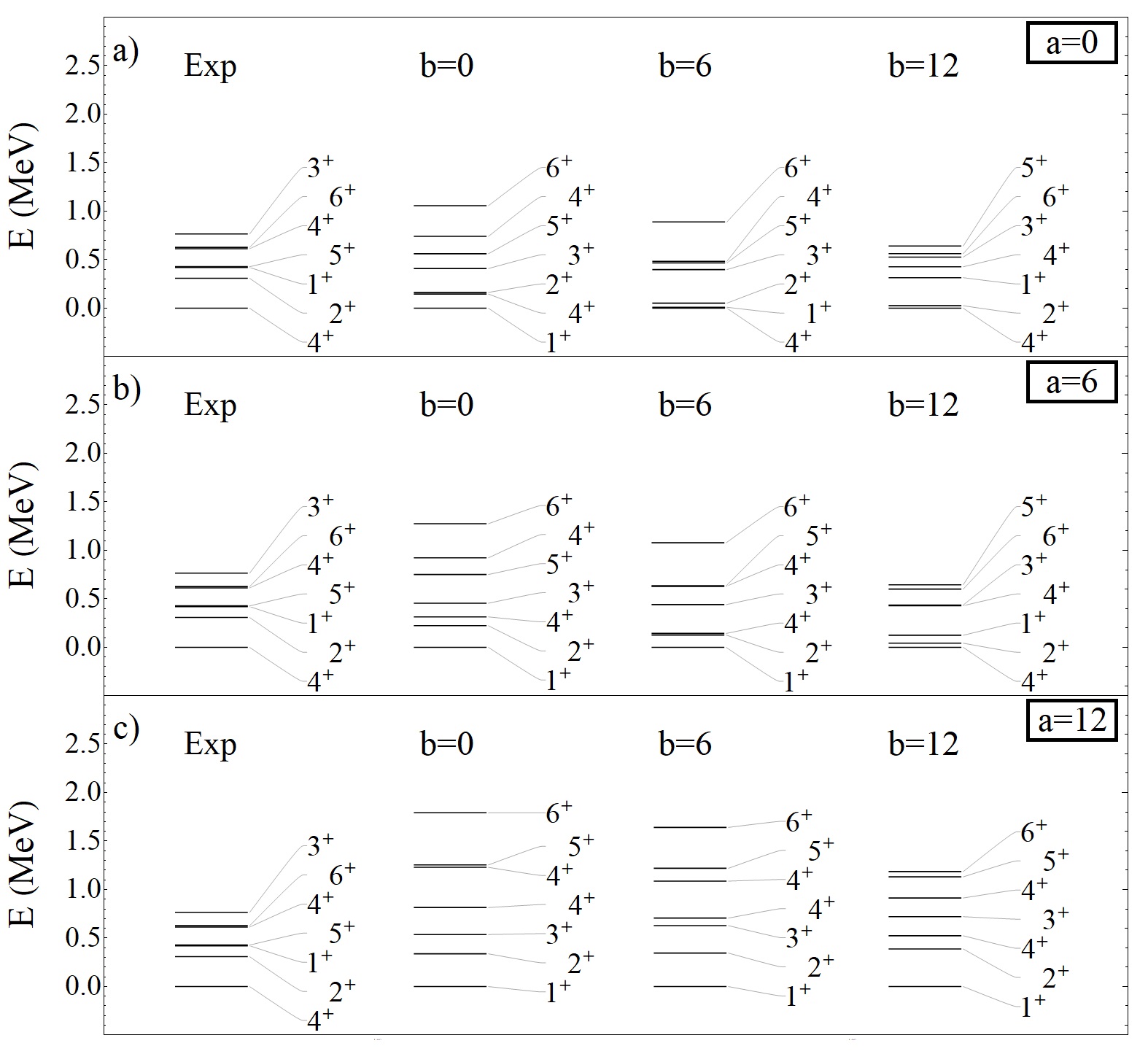}
\caption{Energy spectra of $^{48}V$ as a function of the isovector $b$ and isoscalar $a$  pairing strengths in comparison with experimental data. Results are shown for $a,~b~=0,~6$, and $12$.}
\end{figure}

The first point to note is that the experimental ground state of this nucleus has spin and parity $4^+$. The model is able to reproduce the $4^+$ ground state for the optimal set of parameters $a=0$ and $b=6$. If $a$ is increased, however, it no longer produces the correct ground state spin and parity, confirming again that it is critical to set $a=0$ to suppress erroneous features in the low-energy spectrum when treating nuclei with a neutron excess.

Results for the energy spectra of  $^{48}Ti$ are exhibited in Fig. 15 as a function of the parameter $b$. While the model with the optimal value of $b=6$ reproduces the experimental spectrum reasonably well, the agreement is rapidly lost when we increase or decrease $b$.

\begin{figure}[hbtp]
\centering
\includegraphics[width=1.0\textwidth]{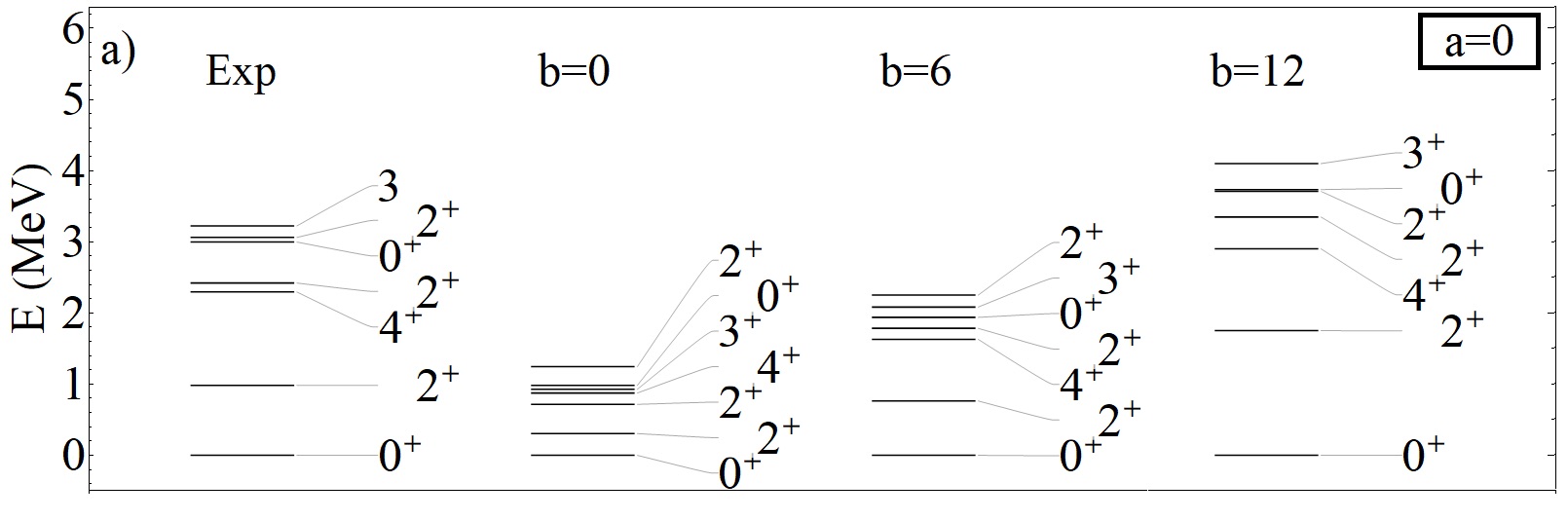}
\caption{Energy spectra of $^{48}Ti$  as a function of the isovector $b$ pairing strength in comparison with the experimental data. Results are shown for $a = 0,~b~=0,~6$, and  $12$.}
\end{figure}

Finally we show results in Fig. 16 for the states of the ground state rotational band in the $N=Z$ nucleus $^{48}Cr$. These results were already discussed in Section 3.2.4 in the main body of the paper.

\begin{figure}[hbtp]
\centering
\includegraphics[width=1.0\textwidth]{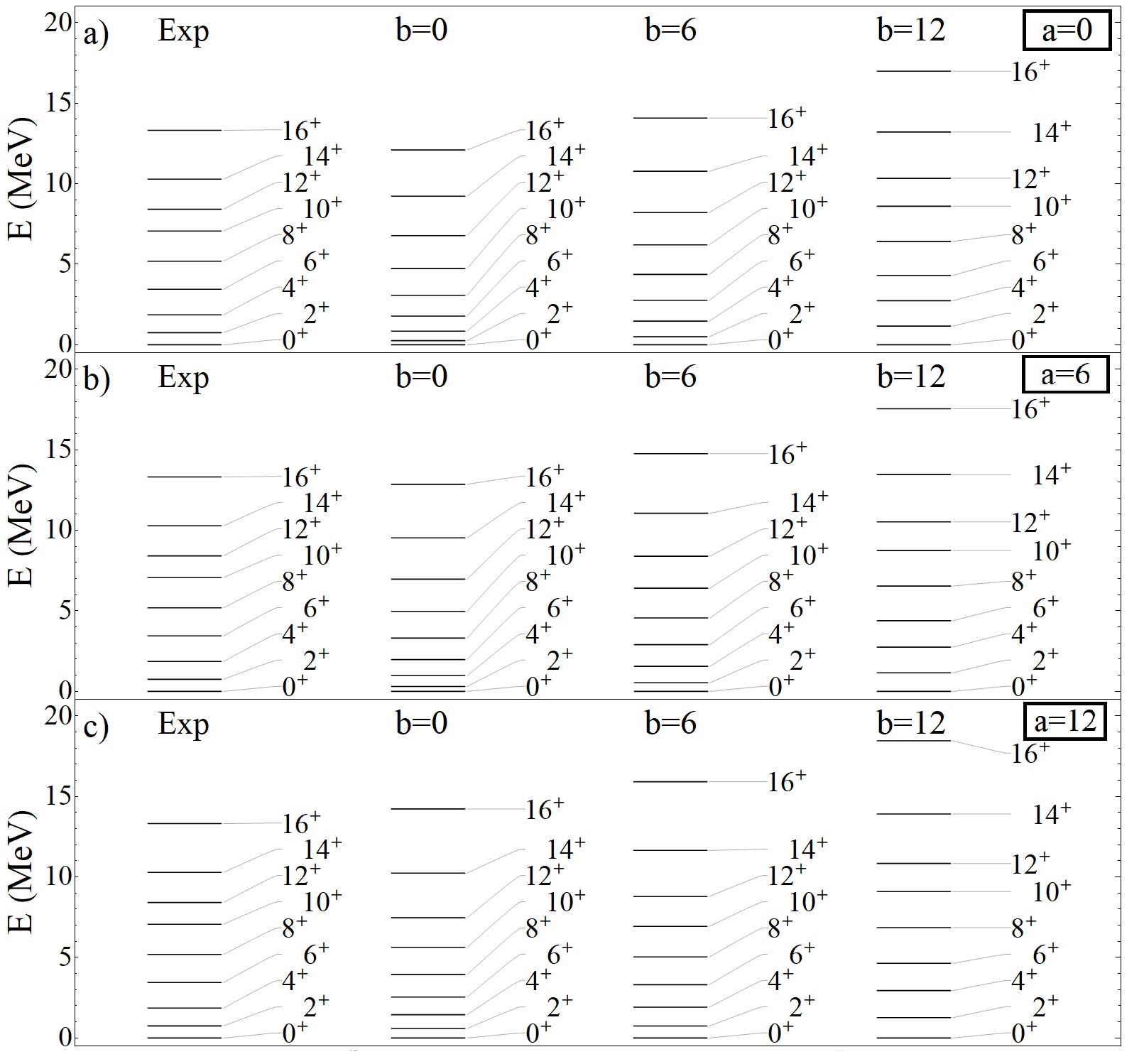}
\caption{Rotational band of $^{48}Cr$ as a function of the isovector $b$ and isoscalar $a$  pairing strengths in comparison with the experimental data. Results are shown for $a,~b~=0,~6$, and $12$.}
\end{figure}


\


\begin{thebibliography}{breitestes Label}
\bibitem{Caurier1995}
Missing and quenched Gamow-Teller strength,
E. Caurier, A. Poves, A.P. Zuker, Phys. Rev.  Lett. 74 (1995) 1517-1520.
%
\bibitem{Cole2012}
Gamow-Teller strengths and electron-capture rates for pf-shell nuclei of relevance for late stellar evolution,
A. L. Cole, T. S. Anderson, R. G. T. Zegers, Sam M. Austin, B. A. Brown, L. Valdez, S. Gupta, G. W. Hitt, and O. Fawwaz,
Phys. Rev. C 86 (2012) 015809.
%
\bibitem{Kumar2016}
Shell model description of Gamow-Teller strengths in pf-shell nuclei,
Vikas Kumar and P.C. Srivastava, Eur. Phys. J. A 52 (2016) 181.
%
\bibitem{Kumar2016b}
Nuclear beta-decay half-lives for fp and fpg shell nuclei,
Vikas Kumar, P C Srivastava and Hantao Li, J. Phys. G: Nucl. Part. Phys. 43 (2016) 105104.
%
\bibitem{Langanke2003}
Nuclear weak-interaction processes in stars,
K. Langanke and G. Martínez-Pinedo,
Rev. Mod. Phys. 75 (2003) 819.
%
\bibitem{Balasi2015}
Neutrino-nucleus reactions and their role for supernova dynamics and nucleosynthesis,
K.G. Balasi, K. Langanke, G. Martínez-Pinedo,
Prog. Part. Nucl. Phys. 85 (2015) 33-81.
%
\bibitem{Saakyan2013}
Two-Neutrino Double-Beta Decay,
Ruben Saakyan, Annu. Rev. Nucl. Part. Sci.  63 (2013) 503-529.
%
\bibitem{Dolinsky2019}
Neutrinoless Double-Beta Decay: Status and Prospects,
Michelle J. Dolinski, Alan W.P. Poon, and Werner Rodejohann, Annu. Rev. Nucl. Part. Sci.  69 (2019) 219-251.
%
\bibitem{Heger2005} Neutrino Nucleosynthesis,
A. Heger, E. Kolbe, W. C. Haxton, K. Langanke, G. Mart\'\i ­nez-Pinedo, and S. E.Woosley, Phys. Lett. B 606  (2005) 258.
%
\bibitem{Suzuki2013} Element synthesis in the supernova environment and neutrino oscillations, T. Suzuki and T. Kajino, J. Phys. G 40 (2013) 083101.
%
\bibitem{Byelikov2007}
Gamow-Teller Strength in the Exotic Odd-Odd Nuclei $^{138}$La and $^{180}$Ta and Its Relevance for Neutrino Nucleosynthesis,
A. Byelikov, T. Adachi, H. Fujita, K. Fujita, Y. Fujita, K.
Hatanaka, A. Heger, Y. Kalmykov, K. Kawase, K. Langanke et al., Phys. Rev. Lett. 98  (2007) 082501.
%
\bibitem{Goodman1979}
Hartree-Fock-Bogoliubov Theory with Applications to Nuclei, A.L. Goodman,  in : Advances in Nuclear Physics, Vol. 11, eds. J.W. Negele and E. Vogt (Plenum, New York, 1979) 263.
%
\bibitem{Doba04} For a concise historical revision with references see J. Dobaczewski (2004), https://www.fuw.edu.pl/~dobaczew/nppair60w/node2.html.
%
\bibitem{Lei2011}
Systematic study of proton-neutron pairing correlations in the nuclear shell model,
Y. Lei, S. Pittel, N. Sandulescu, A. Poves, B. Thakur and Y. M. Zhao, Phys. Rev. C 84 (2011) 044318.
%
\bibitem{Petermann2007}
Breaking of the SU(4) limit for the Gamow-Teller strength in N ~ Z nuclei,
I. Petermann, G. Martinez-Pinedo, K. Langanke, and E. Caurier, Eur. Phys. J. A 34 (2007) 319-324.
%
\bibitem{Lerma2007}
Exact Solution of the Spin-Isospin Proton-Neutron Pairing Hamiltonian,
S. Lerma H., B. Errea, J. Dukelsky, and W. Satula,
Phys. Rev. Lett. 99 (2007) 032501.
%
\bibitem{Sandulescu2009}
Isovector neutron-proton pairing with particle number projected BCS,
N. Sandulescu, B. Errea, and J. Dukelsky,
Phys. Rev. C 80 (2009) 044335.
%
\bibitem{Dobes1998}
Boson mappings and four-particle correlations in algebraic neutron-proton pairing models,
J. Dobes, S. Pittel, Phys. Rev. C57 (1998) 688.
%
\bibitem{Dobaczewski} Symmetry restoration in the mean-field description of proton-neutron pairing,
A.M. Romero, J. Dobaczewski and A. Pastore, Phys. Lett. B 795 (2019)177.
%
\bibitem{Lenske2019}
Heavy ion charge exchange reactions as probes for nuclear beta -decay,
Horst Lenske, Francesco Cappuzzello, Manuela Cavallaro, Maria Colonna, Prog. Part. Nucl. Phys. 109 (2019) 103716
%
\bibitem{Ichimurai2006}
Spin-isospin responses via (p,n) and (n,p) reactions,
M. Ichimura, H. Sakai, T. Wakasa,
Prog. Part. Nucl. Phys. 56 (2006) 446-531.
%
\bibitem{Frekers2018}
Charge-exchange reactions and the quest for resolution,
D. Frekers and M. Alanssari, Eur. Phys. J. A  54 (2018)177.
%
\bibitem{Amos2007}
Charge-exchange reaction cross sections and the Gamow-Teller strength for double beta decay,
K. Amos, Amand Faessler and V. Rodin, Phys. Rev. C 76  (2007) 014604.
%
\bibitem{Sarriguren2001}
Beta decay in odd A and even even proton rich Kr isotopes,
P. Sarriguren, E. Moya de Guerra, A. Escuderos, Phys. Rev. C 64 (2001) 064306.
%
\bibitem{Moreno2006}
Gamow-Teller strength distributions in Xe isotopes,
O. Moreno, R. Alvarez-Rodriguez, P. Sarriguren, E. Moya de Guerra, J.M. Udias et al.
Phys. Rev. C 74 (2006) 054308.
%
\bibitem{Cha83}
$\sigma \tau^+$ strength in nuclei,
D. Cha, Phys. Rev. C 27(1983) 2269.
%
\bibitem{Nabi2013}
Comparison of Gamow-Teller strengths in the random phase approximation,
Jameel-Un Nabi, Calvin W. Johnson,
J. Phys. G 40 (2013) 065202.
%
\bibitem{Niu2016}
Quasiparticle random-phase approximation with quasiparticle-vibration coupling: Application to the Gamow-Teller response of the superfluid nucleus $^{120}$ Sn,
Y.F. Niu, G. Colo, E. Vigezzi, C.L. Bai, H. Sagawa,
Phys.Rev.C 94 (2016) 064328.
%
\bibitem{Sarriguren2018}
Beta-decay properties of neutron-rich Ca, Ti, and Cr isotopes,
P. Sarriguren, A. Algora, and G. Kiss, Phys. Rev. C 98 (2018) 024311.
%
\bibitem{Osterfeld1992}
Nuclear spin and isospin excitations,
Franz Osterfeld, Rev. Mod. Phys. 64 (1992) 491-558.
%
\bibitem{Osterfeld1991}
Nuclear spin and isospin excitations
Franz Osterfeld, Rev. Mod. Phys. 64 (1992) 491-558.
%
\bibitem{Suhonen2013}
Probing the quenching of $g_A$ by single and double beta decays,
Jouni Suhonen, Osvaldo Civitarese, Phys. Lett. B 725 (2013) 153-157.
%
\bibitem{Suhonen2017}
Value of the Axial-Vector Coupling Strength in $\beta$ and $\beta\beta$ Decays: A Review,
Jouni T. Suhonen, Front.in Phys. 5 (2017) 55.
%
\bibitem{Hirsch1990}
Gamow-Teller strength functions and two-neutrino double-beta decay,
J Hirsch, E Bauer, F Krmpotic,
Nucl. Phys. A 516 (2), 304-324.
%
\bibitem{Udagawa1994}
Delta excitations in nuclei and their decay properties,
T. Udagawa, P. Oltmanns, F. Osterfeld, S.W. Hong, Phys. Rev. C 49 (1994) 3162-3181.
%
\bibitem{Cattapan2002}
The role of the Delta in nuclear physics,
G. Cattapan, L.S. Ferreira, Phys. Rep. 362 (2002) 303-407.
%
\bibitem{Caurier1995b}
Gamow-Teller strength in Fe-54 and Fe-56,
E. Caurier, G. Martinez-Pinedo, A. Poves, A.P. Zuker, Phys. Rev. C 52 (1995) R1736-R1740.
%
\bibitem{Gysbers2019}
Discrepancy between experimental and theoretical $\beta$-decay rates resolved from first principles,
Gysbers, P., Hagen, G., Holt, J.D. et al.,  Nat. Phys. 15, 428?431 (2019).
%
\bibitem{Hirsch1988}
On the 2p-2h excitations and the quenching of the Gamow-Teller strength,
J. Hirsch, A. Mariano, M. Faig, F. Krmpotik,
Phys. Lett. B 210 (1988) 55-60.
%
\bibitem{Bai2009}
Quenching of Gamow-Teller strength due to tensor correlations in 90Zr and 208Pb,
C.L. Bai, H.Q. Zhang, X.Z. Zhang, F.R. Xu, H. Sagawa et al., Phys. Rev. C 79 (2009) 041301.
%
\bibitem{Marketin2012}
Role of momentum transfer in the quenching of Gamow-Teller strength,
T. Marketin, G. Martinez-Pinedo, N. Paar, D. Vretenar, Phys. Rev. C 85 (2012) 054313.
%
\bibitem{Wang2018b}
Quenching of nuclear matrix elements for $0\nu \beta\beta$ decay by chiral two-body currents,
Long-Jun Wang, Jonathan Engel, and Jiang Ming Yao,
Phys. Rev. C 98 (2018) 031301(R).
%
\bibitem{Wang2018}
Shell-model method for Gamow-Teller transitions in heavy deformed odd-mass nuclei,
Long-Jun Wang, Yang Sun, and Surja K. Ghorui, Phys. Rev. C 97 (2018) 044302.
%
\bibitem{Rolfs1988}
Cauldrons in the Cosmos, C.E. Rolfs, W. Rodney (University of Chicago Press, 1988).
%
\bibitem{Bertulani2016}
Frontiers in Nuclear Astrophysics,
C.A. Bertulani, T. Kajino, Prog. Part. Nucl. Phys. 89 (2016) 56-100.
%
\bibitem{Juodagalvis2010}
Improved estimate of electron capture rates on nuclei during stellar core collapse,
A. Juodagalvis, K. Langanke, W.R. Hix, G. Martínez-Pinedo, J.M. Sampaio, Nucl. Phys. A 848 (2010) 454-478.
%
\bibitem{Ejiri2000}
Nuclear spin isospin responses for low-energy neutrinos,
H. Ejiri, Phys. Rep. 338 (2000) 265-351.
%
\bibitem{Ejiri2019}
Neutrino-nuclear responses for astro-neutrinos, single beta decays and double beta decays,
H. Ejiri, J. Suhonen, K. Zuber, Phys. Rep. 797 (2019) 1-102.
%
\bibitem{Langanke2001}
Unblocking of the Gamow-Teller strength in stellar electron capture on neutron-rich germanium isotopes,
K. Langanke, E. Kolbe, and D. J. Dean, Phys. Rev. C 63 (2001) 032801R.
%
\bibitem{Tomoda1991}
Double beta decay,
T. Tomoda. Rept. Prog. Phys. 54 (1991) 53-126.
%
\bibitem{Faessler1998}
Double beta decay,
Amand Faessler, Fedor Simkovic, J. Phys. G 24 (1998) 2139-2178.
%
\bibitem{Suhonen1998}
Weak-interaction and nuclear-structure aspects of nuclear double beta decay,
J Suhonen, O Civitarese, Phys. Rep. 300 (1998) 123-214.
%
\bibitem{Vergados2002}
The Neutrinoless double beta decay from a modern perspective,
J.D. Vergados, Phys. Rep. 361 (2002) 1-56.
%
\bibitem{Zdesenko2003}
The future of double beta decay research,
Yu. Zdesenko, Rev. Mod. Phys. 74 (2003) 663-684.
%
\bibitem{Elliot2004}
Double beta decay,
Steven R. Elliott, Jonathan Engel, J. Phys. G 30 (2004) R183-R215.
%
\bibitem{Avignone2008}
Double Beta Decay, Majorana Neutrinos, and Neutrino Mass,
Frank T. III Avignone, Steven R. Elliott, Jonathan Engel, Rev. Mod. Phys. 80 (2008) 481-516.
%
\bibitem{Barabash2011}
Double Beta Decay: Historical Review of 75 Years of Research,
A.S. Barabash, Phys. Atom. Nucl. 74 (2011) 603-613.
%
\bibitem{Vergados2012}
Theory of Neutrinoless Double Beta Decay,
J.D. Vergados, H. Ejiri, F. Simkovic, Rep. Prog. Phys. 75 (2012) 106301.
%
\bibitem{Bilenky2015}
Neutrinoless Double-Beta Decay: a Probe of Physics Beyond the Standard Model,
S.M. Bilenky, C. Giunti, Int. J. Mod. Phys. A 30 (2015) 04n05, 1530001.
%
\bibitem{Engel2017}
Status and Future of Nuclear Matrix Elements for Neutrinoless Double-Beta Decay: A Review,
Jonathan Engel, Javier Menéndez, Rep. Prog. Phys. 80 (2017) 046301.
%
\bibitem{Suhonen1998}
Weak-interaction and nuclear-structure aspects of nuclear double beta decay,
J Suhonen, O Civitarese, Phys. Rep. 300 (1998) 123-214.
%
\bibitem{Zhao1990}
Shell model calculation for two neutrino double beta decay of Ca-48,
L. Zhao, B.Alex Brown, W.A. Richter, Phys. Rev. C 42 (1990) 1120-1125.
%
\bibitem{Caurier1990}
A Full 0 h-bar omega description of the 2 neutrino beta beta decay of Ca-48,
E. Caurier, A.P. Zuker, A. Poves, Phys. Lett.B 252 (1990) 13-17.
%
\bibitem{Retamosa1995}
Neutrinoless double beta decay of Ca-48,
J. Retamosa, E. Caurier, F. Nowacki, Phys. Rev. C 51 (1995) 371-378.
%
\bibitem{Horoi2007}
Shell-model calculations of two-neutrino double-beta decay rates of Ca-48 with GXPF1A interaction,
M. Horoi, S. Stoica, B.Alex Brown, Phys. Rev. C 75 (2007) 034303.
%
\bibitem{Horoi2010}
Shell Model Analysis of the Neutrinoless Double Beta Decay of Ca-48,
Mihai Horoi, Sabin Stoica, Phys. Rev. C 81 (2010) 024321.
%
\bibitem{Caurier1996}
Shell Model Studies of the Double Beta Decays of Ge-76, Se-82, and Xe-136,
E. Caurier, F. Nowacki, A. Poves, J. Retamosa, Phys. Rev. Lett. 77 (1996) 1954-1957.
%
\bibitem{Horoi2013}
Shell-Model Analysis of the 136 Xe Double Beta Decay Nuclear Matrix Elements,
M. Horoi, B.A. Brown, Phys. Rev. Lett. 110 (2013) 222502.
%
\bibitem{Iwata2016}
Large-Scale Shell-Model Analysis of the Neutrinoless double beta Decay of 48Ca,
 Y. Iwata, N. Shimizu, T. Otsuka, Y. Utsuno, J. Menéndez, M. Honma and T. Abe, Phys. Rev. Lett. 116 (2016) 112502.
%
\bibitem{Shimizu2018}
Double Gamow-Teller Transitions and its Relation to Neutrinoless Double Beta Decay,
Noritaka Shimizu, Javier Menendez, and Kentaro Yako, Phys. Rev. Lett. 120 (2018) 142502.
%
\bibitem{Koonin1997}
Shell model Monte Carlo methods,
S.E. Koonin, D.J. Dean, K. Langanke, Phys. Rep. 278 (1997) 1-77.
%
\bibitem{Vogel1986}
Suppression of the two-neutrino double-beta decay by nuclear-structure effects,
P. Vogel and M. R. Zirnbauer,
Phys. Rev. Lett. 57 (1986) 3148.
%
\bibitem{Engel1988}
Nuclear structure effects in double-beta decay,
J. Engel, P. Vogel, and M. R. Zirnbauer,
Phys. Rev. C 37 (1988) 731.
%
\bibitem{Muto1988}
Calculation of 2$\nu$ double beta decay of 76 Ge, 82 Se, 128,130 Te,
K. Muto, H.V. Klapdor, Phys.Lett.B 201 (1988) 420-424.
%
\bibitem{Hirsch1990}
Reconstruction of isospin and spin - isospin symmetries and double beta decay,
J. Hirsch, F. Krmpotic, Phys.Rev.C 41 (1990) 792-795.
%
\bibitem{Civitarese1987}
Suppression of the Two Neutrino Double Beta Decay,
O. Civitarese, A. Faessler, T. Tomoda, Phys. Lett. B194 ( 1987) 11-14.
%
\bibitem{Staudt1990}
$\beta \beta$ decay of Ge-76 with renormalized effective interaction derived from Paris, Bonn and Reid potentials,
A. Staudt, T.T.S. Kuo, H.V. Klapdor- Kleingrothaus, Phys. Lett. B 242 (1990) 17-23.
%
\bibitem{Pirinen2015}
Systematic approach to $\beta$ and $2\beta$ decays of mass A=100-136 nuclei,
Pekka Pirinen, Jouni Suhonen, Phys. Rev. C 91 (2015) 054309.
%
\bibitem{Engel1989}
Double Beta Decay in the Generalized Seniority Scheme,
J. Engel, P. Vogel, Xiang-Dong Ji, S. Pittel, Phys. Lett. B 225 (1989) 5-9.
%
\bibitem{Civitarese1990}
Suppression of the two neutrino $\beta \beta$ decay: Particle number projected results,
O. Civitarese, Amand Faessler, J. Suhonen, X.R. Wu,
Phys.Lett.B 251 (1990) 333-337;
Suppression of the two neutrino beta beta decay in a particle number projected quasiparticle random phase approximation,
O. Civitarese, Amand Faessler, J. Suhonen, X.R. Wu, Nucl. Phys. A 524 (1991) 404-424.
%
\bibitem{Suhonen1991}
The Neutrinoless double beta decay of Ge-76, Se-82, Kr-86, Cd-114, Te-128, Te-130 and Xe-134, Xe-136 in the framework of a relativistic quark confinement model,
J. Suhonen, S.B. Khadkikar, Amand Faessler, Nucl. Phys. A 535 (1991) 509-547.
%
\bibitem{Suhonen2011}
On the double-beta decays of Zn-70, Kr-86, Zr-94, Ru-104, Pd-110 and Sn-124,
Jouni Suhonen, Nucl. Phys. A 864 (2011) 63-90.
%
\bibitem{Suhonen2012}
Review of the properties of the $0 \nu \beta^- \beta^-$ nuclear matrix elements,
Jouni Suhonen, Osvaldo Civitarese, J. Phys. G 39 (2012) 124005.
%
\bibitem{Simkovic2013}
$0 \nu \beta \beta$  and $2 \nu \beta \beta$ nuclear matrix elements, quasiparticle random-phase approximation, and isospin symmetry restoration,
Fedor Simkovic, Vadim Rodin, Amand Faessler, Petr Vogel, Phys.Rev.C 87 (2013) 045501.
%
\bibitem{Griffiths1992}
Double beta decay to excited $0^+$ states: Decay of Mo-100,
A. Griffiths, P. Vogel, Phys.Rev.C 46 (1992) 181-187.
%
\bibitem{Suhonen1993}
Two neutrino beta beta decay to excited states: The $0^+ \rightarrow 2^+$ decay of Xe-136,
J. Suhonen, O. Civitarese, Phys. Lett. B 308 (1993) 212-215.
%
\bibitem{Suhonen1994}
Quasiparticle random phase approximation analysis of the double beta decay of Mo-100 to the ground state and excited states of Ru-100,
J. Suhonen, O. Civitarese, Phys. Rev. C 49 (1994) 3055-3060.
%
\bibitem{Civitarese1994}
Two neutrino double beta decay to excited one and two phonon states,
O. Civitarese, J. Suhonen, Nucl. Phys. A 575 (1994) 251-268.
%
\bibitem{Aunola1996}
Systematic study of beta and double beta decay to excited final states,
M. Aunola, J. Suhonen, Nucl. Phys. A 602 (1996) 133-166.
%
\bibitem{Cheuon1993}
Two neutrino double beta decay in coupled QRPA with neutron proton pairing,
M.K. Cheoun, A. Bobyk, A. Faessler, F. Simkovic, G. Teneva, Nucl. Phys. A 564 (1993) 329-344.
%
\bibitem{Toivanen1995}
Renormalized proton neutron quasiparticle random phase approximation and its application to double beta decay,
J. Toivanen, J. Suhonen, Phys. Rev. Lett. 75 (1995) 410-413.
%
\bibitem{Schwieger1996}
The Pauli principle, QRPA and the two neutrino double beta decay,
J. Schwieger, F. Simkovic, Amand Faessler, Nucl.Phys.A 600 (1996) 179-192
%
\bibitem{Toivanen1997}
Study of several double-beta-decaying nuclei using the renormalized proton neutron quasiparticle random-phase approximation
J. Toivanen, J. Suhonen, Phys. Rev. C 55 (1997) 2314-2323.
%
\bibitem{Hirsch1996}
Renormalized QRPA and double beta decay: A Critical analysis,
Jorge G. Hirsch, Peter O. Hess, Osvaldo Civitarese, Phys. Rev. C 54 (1996) 1976-1981.
%
\bibitem{Engel1997}
Neutron-proton correlations in an exactly solvable model,
J. Engel, S. Pittel, M. Stoitsov, P. Vogel, J. Dukelsky, Phys. Rev. C 55 (1997) 1781-1788.
%
\bibitem{Hirsch1997}
Single and double beta decay Fermi transitions in an exactly solvable model,
Jorge G. Hirsch, Peter O. Hess, Osvaldo Civitarese, Phys. Rev. C 56 (1997) 199.
%
\bibitem{Stoica2001}
Critical view on double-beta decay matrix elements within quasi random phase approximation-based methods,
S. Stoica, H.V. Klapdor-Kleingrothaus, Nucl. Phys. A 694 (2001) 269-294.
%
\bibitem{Rodin2003}
On the uncertainty in the $0 \nu \beta \beta$ decay nuclear matrix elements,
V.A. Rodin, Amand Faessler, F. Simkovic, Petr Vogel, Phys. Rev. C 68 (2003) 044302.
%
\bibitem{Rodin2006}
Assessment of uncertainties in QRPA $0\nu \beta\beta$-decay nuclear matrix elements,
V.A. Rodin, A. Faessler, F. Simkovic, P. Vogel, Nucl. Phys. A 766 (2006) 107-131, Nucl. Phys. A 793 (2007) 213-215 (erratum).
%
\bibitem{Raduta1993}
Unified description of the  $2 \nu \beta \beta$  decay in spherical and deformed nuclei,
A.A. Raduta, A. Faessler, D.S. Delion, Nucl. Phys. A 564 (1993) 185-203.
%
\bibitem{Simkovic2004}
Two neutrino double beta decay of Ge-76 within deformed GRPA: A new suppression mechanism,
Fedor Simkovic, Larisa Pacearescu, Amand Faessler, Nucl. Phys. A 733 (2004) 321-350.
%
\bibitem{Alvarez2004}
A Deformed QRPA formalism for single and two-neutrino double beta decay,
R. Alvarez-Rodriguez, P. Sarriguren, E. Moya de Guerra, L. Pacearescu, Amand Faessler, Phys. Rev. C 70 (2004) 064309.
%
\bibitem{Yousef2008}
Two-neutrino double beta decay of deformed nuclei within QRPA with realistic interaction,
Mohamed Saleh Yousef, Vadim Rodin, Amand Faessler, Fedor Simkovic, Phys. Rev. C 79 (2009) 014314
%
\bibitem{Castanos1994}
Double-beta decay in the pseudo SU(3) scheme,
Octavio Casta\~nos, Jorge G. Hirsch, Osvaldo Civitarese, Peter O. Hess, Nucl. Phys. A 571 (1994) 276-300,
%
\bibitem{Hirsch1994}
Neutrinoless double beta decay in heavy deformed nuclei,
Jorge G. Hirsch, O. Casta\~nos, P.O. Hess, Nucl. Phys. A 582 (1995) 124-140.
%
\bibitem{Singh2007}
Nuclear deformation and the two neutrino double-beta decay in Xe-124,126, Te-128,130, Ba-130-132 and Nd-150 isotopes,
S. Singh, R. Chandra, P.K. Rath, P.K. Raina, J.G. Hirsch, Eur. Phys. J. A 33 (2007) 375-388.
%
\bibitem{Chaturvedi2008}
Nuclear deformation and neutrinoless double-beta decay of Zr-94, Zr-96, Mo-98, Mo-100, Ru-104, Pd-110, Te-128, Te-130, and Nd-150 nuclei within a mechanism involving neutrino mass,
K. Chaturvedi, R. Chandra, P.K. Rath, P.K. Raina(, J.G. Hirsch, Phys. Rev. C 78 (2008) 054302.
%
\bibitem{Rath2010}
Uncertainties in nuclear transition matrix elements for neutrinoless  $\beta \beta$ decay within the PHFB model,
P.K. Rath, R. Chandra, K. Chaturvedi, P.K. Raina, J.G. Hirsch, Phys. Rev. C 82 (2010) 064310.
%
\bibitem{Rath2013}
Neutrinoless  $beta \beta$  decay transition matrix elements within mechanisms involving light Majorana neutrinos, classical Majorons, and sterile neutrinos,
P.K. Rath, R. Chandra, K. Chaturvedi, P. Lohani, P.K. Raina, Phys. Rev. C 88 (2013) 6, 064322.
%
\bibitem{Rath2019}
Nuclear Transition Matrix Elements for Double-beta Decay Within PHFB Model,
P.K. Rath, R. Chandra, K. Chaturvedi and P.K. Raina, Front.  Phys. 7 (2019) 64.
%
\bibitem{NUMEN2018}
The NUMEN project: NUclear Matrix Elements for Neutrinoless double beta decay,
F. Cappuzzello, C. Agodi, M. Cavallaro, D. Carbone, S. Tudisco et al.,
Eur. Phys. J. A 54 (2018) 5, 72.
%
\bibitem{Dean2003}
Pairing in nuclear systems: from neutron stars to finite nuclei,
D. J. Dean, M. Hjorth-Jensen, Rev. Mod. Phys. 75 (2003) 607.
%
\bibitem{BCS1957}
Theory of Superconductivity,
J. Bardeen, L.N. Cooper, and J.R. Schrieffer, Phys. Rev. 108 (1957) 1175.
%
\bibitem{Bohr1958}
Possible Analogy between the Excitation Spectra of Nuclei and Those of the Superconducting Metallic State,
A. Bohr, B.R. Mottelson, and D. Pines, Phys. Rev. 110 (1958) 936.
%
\bibitem{Kerman1961}
Accuracy of the Superconductivity Approximation for Pairing Forces in Nuclei,
A.K. Kerman, R.D. Lawson, M.H. Macfarlane, Phys. Rev. 124 (1961) 162-167.
%
\bibitem{Nogami1964}
Improved Superconductivity Approximation for the Pairing Interaction in Nuclei,
Yukihisa Nogami, Phys. Rev. 134 (1964) B313-B321.
%
\bibitem{Dietrich1964}
Conservation of Particle Number in the Nuclear Pairing Model,
K. Dietrich, H.J. Mang, and J.H. Pradal, Phys. Rev. 135 (1964) B22.
%
\bibitem{Richardson1963}
A Restricted Class of Exact Eigenstates of the Pairing- Force Hamiltonian,
R. W. Richardson, Phys. Lett. 3 (1963) 277-279;
Exact eigenstates of the pairing-force Hamiltonian,
R.W.Richardson, N.Sherman, Nucl. Phys. 52 (1964) 221-238.
%
\bibitem{Dukelsky2004}
Exactly solvable Richardson-Gaudin models for many-body quantum systems,
J. Dukelsky, S. Pittel, and G. Sierra, Rev. Mod. Phys. 76, 643-662 (2004).
%
\bibitem{Dukelsky2013}
Exact Solutions for Pairing Interactions,
J. Dukelsky and S. Pittel, in Fifty Years of Nuclear BCS, Pairing in Finite Systems,
Edited By: Ricardo A Broglia and Vladimir Zelevinsky, World Scientific, Singapore (2013), pp. 200-211.
%
\bibitem{Ortiz2005}
Exactly-solvable models derived from a generalized Gaudin algebra,
G. Ortiz, R. Somma, J. Dukelsky, and S. Rombouts,  Nucl. Phys. B 707 (2005) 421-457.
%
\bibitem{Afanasjev2013}
Isoscalar and Isovector Neutron-Proton Pairing,
A. V. Afanasjev, in Fifty Years of Nuclear BCS, Pairing in Finite Systems,
Edited By: Ricardo A Broglia and Vladimir Zelevinsky, World Scientific, Singapore (2013), pp. 138-153.
%
\bibitem{Frauendorf2014}
Overview of neutron-proton pairing,
S. Frauendorf, A.O. Macchiavelli,
Prog. Part. Nucl. Phys. 78 (2014) 24-90.
%
\bibitem{Negrea2018}
Isovector and isoscalar proton-neutron pairing in N$>$Z nuclei,
D. Negrea, P. Buganu, D. Gambacurta, N. Sandulescu,
Phys. Rev. C 98 (2018) 064319.
%
\bibitem{Baranger1960}
Extension of the Shell Model for Heavy Spherical Nuclei,
Michel Baranger, Phys. Rev. 120 (1960) 3, 957.
%
\bibitem{Kumar1968}
Nuclear deformations in the pairing-plus-quadrupole model,
Krishna Kumar, Michel Baranger, Nucl.Phys.A 110 (1968) 529-554;
Michel Baranger, Krishna Kumar, Nucl.Phys.A 122 (1968) 241-272, Nucl.Phys.A 141 (1970) 674-674 (erratum);
Nuclear deformations in the pairing-plus-quadropole model (II). Discussion of the validity of the model,
Michel Baranger, Krishna Kumar, Nucl.Phys.A 110 (1968) 490-528.
%
\bibitem{Bes1969}
The pairing-plus-quadrupole model,
Daniel R. Bes and Raymond A. Sorensen, M. Baranger et al. (eds.), Advances in Nuclear Physics, Plenum Press 1969, 129.
%
\bibitem{Brack1972}
Funny Hills: The Shell-Correction Approach to Nuclear Shell Effects and Its Applications to the Fission Process,
M. Brack, Jens Damgaard, A.S. Jensen, H.C. Pauli, V.M. Strutinsky et al., Rev. Mod. Phys. 44 (1972) 320-405.
%
\bibitem{Kishimoto1976}
Description of nuclear collective motions in terms of the boson expansion technique,
T. Kishimoto, T. Tamura, Nucl. Phys. A 270 (1976) 317-380.
%
\bibitem{Arima1978}
Interacting boson model of collective states. 1. The Vibrational limit,
A. Arima, F. Iachello, Annals Phys. 99 (1976) 253-317;
Interacting boson model of collective nuclear states. II. The rotational limit,
A. Arima, F. Iachello, Annals Phys. 111 (1978) 201-238;
New Symmetry in the sd Boson Model of Nuclei: The Group O(6);
A. Arima, F. Iachello, Phys.Rev.Lett. 40 (1978) 385;
Interacting boson model of collective nuclear states. 4. The O(6) limit,
A. Arima, F. Iachello, Annals Phys. 123 (1979) 468.
%
\bibitem{Bijker1980}
Description of the Pt and Os isotopes in the interacting boson model,
R. Bijker, A.E.L. Dieperink, O. Scholten, R. Spanhoff, Nucl. Phys. A 344 (1980) 207-232.
%
\bibitem{Isacker1981}
Classical limit of the interacting boson Hamiltonian,
P. Van Isacker, Jin-Quan Chen, Phys. Rev. C 24 (1981) 684-689.
%
\bibitem{Klein1991}
Boson realizations of Lie algebras with applications to nuclear physics,
Abraham Klein, E.R. Marshalek, Rev. Mod. Phys. 63 (1991) 375-558.
%
\bibitem{Dufour1996}
The realistic collective nuclear Hamiltonian,
Marianne Dufour), Andres Zuker, Phys. Rev. C 54 (1996) 1641-1660.
%
\bibitem{Zuker1995}
Spherical shell model description of rotational motion,
A.P. Zuker, J. Retamosa, A. Poves, E. Caurier, Phys. Rev. C 52 (1995) R1741-R1745.
%
\bibitem{Caurier2005}
The shell model as a unified view of nuclear structure,
E. Caurier, G. Martínez-Pinedo, F. Nowacki, A. Poves, and A. P. Zuker,
Rev. Mod. Phys. 77 (2005) 427.
%
\bibitem{Elliot1958}
Collective motion in the nuclear shell model. 1. Classification schemes for states of mixed configuration,
J.P. Elliott, Proc. Roy. Soc. Lond. A A245 (1958) 128;
Collective motion in the nuclear shell model. 2. The Introduction of intrinsic wave functions,
J.P. Elliott, Proc. Roy. Soc. Lond. A A245 (1958) 562-581.
%
\bibitem{Hecht1973}
Symmetries in nuclei,
K.T. Hecht, Ann. Rev. Nucl. Part. Sci. 23 (1973) 123-161.
%
\bibitem{Ratna1973}
Search for a coupling scheme in heavy deformed nuclei: The pseudo SU(3) model,
R.D.Ratna Raju, J.P. Draayer, K.T. Hecht, Nucl. Phys. A 202 (1973) 433-466.
%
\bibitem{Troltenier1995}
Generalized pseudo-SU(3) model and pairing,
D. Troltenier, C. Bahri, J.P. Draayer , Nucl. Phys. A 586 (1995) 53-72.
%
\bibitem{Blokhin1995}
Origin of Pseudospin Symmetry,
A.L. Blokhin, C. Bahri, J.P. Draayer(, Phys. Rev. Lett. 74 (1995) 4149-4152.
%
\bibitem{Meng1998}
Pseudospin symmetry in relativistic mean field theory,
J. Meng, K. Sugawara-Tanabe, S. Yamaji, P. Ring, A. Arima, Phys. Rev. C 58 (1998) R628-R631.
%
\bibitem{Ginocchio2005}
Relativistic symmetries in nuclei and hadrons,
J.N. Ginocchio, Phys. Rep. 414 (2005) 165-261.
%
\bibitem{Liang2015}
Hidden pseudospin and spin symmetries and their origins in atomic nuclei,
Haozhao Liang, Jie Meng, Shan-Gui Zhou, Phys. Rep. 570 (2015) 1-84.
%
\bibitem{Lerma2011}
Isovectorial pairing in solvable and algebraic models,
Sergio Lerma, Carlos E Vargas and Jorge G Hirsch, Journal of Physics: Conference Series 322 (2011) 012011,
%
\bibitem{Hara1995}
Projected Shell Model and High-Spin Spectroscopy,
Kenji Hara, Yang Sun, Int. J. Mod. Phys. E 4 (1995) 637-785.
%
\bibitem{Sun2000}
Multiphonon gamma vibrational bands and the triaxial projected shell model,
Yang Sun, Kenji Hara, Javid A. Sheikh, Jorge G. Hirsch, Victor Velazquez et al., Phys.Rev.C 61 (2000) 064323.
%
\bibitem{Decharge1980}
Hartree-Fock-Bogolyubov calculations with the D1 effective interactions on spherical nuclei,
J. Decharge, D. Gogny, Phys. Rev. C 21 (1980) 1568-1593.
%
\bibitem{Wood1992}
Coexistence in even-mass nuclei,
J.L. Wood, K. Heyde, W. Nazarewicz, M. Huyse, P. van Duppen, Phys. Rep. 215 (1992) 101-201.
%
\bibitem{Poves1981}
Theoretical Spectroscopy and the fp shell,
A, Pover, A. Zuker, Phys. Rep. 70 (1981) 235-314.
%
\bibitem{Honma2005}
Effective interaction for nuclei of A = 50-100 and Gamow-Teller properties,
M Honma, T Otsuka, T Mizusaki, M Hjorth-Jensen and B A Brown, Journal of Physics: Conference Series 20 (2005) 2.
%
\bibitem{Caurier1995c}
Full pf shell model study of A=48 nuclei,
E. Caurier, A.P. Zuker, A. Poves, G. Martinez-Pinedo, Phys. Rev. C 50 (1994) 225-236.
%
\bibitem{Caurier1995d}
Intrinsic vs Laboratory Frame Description of the Deformed Nucleus Cr-48,
E. Caurier, J.L. Egido, G. Martinez-Pinedo, A. Poves, J. Retamosa, Phys. Rev. Lett. 75 (1995) 2466-2469.
%
\bibitem{Nakada1996}
Microscopic description of Gamow-Teller transitions in middle pf shell nuclei by a realistic shell model calculation,
H. Nakada, T. Sebe, J. Phys. G 22 (1996) 1349-1362.
%
\bibitem{Caurier1999}
Shell model calculations of stellar weak interaction rates: I. Gamow-Teller distributions and spectra of nuclei in the mass range A=45-65,
E. Caurier, K. Langanke, G. Martinez-Pinedo, F. Nowacki, Nucl. Phys. A653 (1999) 439-452.
%
\bibitem{Suzuki2011}
Evaluation of electron capture reaction rates in Ni isotopes in stellar environments,
Toshio Suzuki, Michio Honma, Hélène Mao, Takaharu Otsuka, and Toshitaka Kajino,
Phys. Rev. C 83 (2011) 044619.
%
\bibitem{Fujita2005}
Gamow-Teller Strengths in Proton-Rich Exotic Nuclei Deduced in the Combined Analysis of Mirror Transitions,
Y. Fujita, T. Adachi, P. von Brentano, G. P. A. Berg, C. Fransen, D. De Frenne, H. Fujita, K. Fujita, K. Hatanaka, E. Jacobs, K. Nakanishi, A. Negret, N. Pietralla, L. Popescu, B. Rubio, Y. Sakemi, Y. Shimbara, Y. Shimizu, Y. Tameshige, A. Tamii, M. Yosoi, and K. O. Zell,
Phys. Rev. Lett. 95 (2005) 212501.
%
\bibitem{Adachi2006}
High-resolution study of Gamow-Teller transitions from the $^{46}$Ti to the $T_z=0$ nucleus $^{46}$V.
T. Adachi et al., Phys. Rev. C 73 (2006) 024311.
%
\bibitem{Fujita2014}
Observation of Low- and High-Energy Gamow-Teller Phonon Excitations in Nuclei,
Y. Fujita et al.,
Phys. Rev. Lett. 112 (2014) 112502.
%
\bibitem{Molina2015}
$T_z=-1\rightarrow \beta$ decays of$^{54}$Ni, $^{50}$Fe, $^{46}$Cr, and $^{42}$Tt and comparison with mirror $(^3$He,$t)$ measurements
F. Molina et al., Phys. Rev. C 91 (2015) 014301.
%
\bibitem{Yoshida2018}
Systematic shell-model study of $\beta$-decay properties and Gamow-Teller strength distributions in A $\approx$ 40 neutron-rich nuclei,
Sota Yoshida, Yutaka Utsuno, Noritaka Shimizu, and Takaharu Otsuka, Phys. Rev. C 97 (2018) 054321.
%
\bibitem{Martinez1996}
Effective $g_A$ in the pf shell,
G. Martínez-Pinedo, A. Poves, E. Caurier, and A. P. Zuker,
Phys. Rev. C 53 (1996) R2602(R).
%
\bibitem{Auerbach1993}
Correlation between the quenching of total GT+ strength and the increase of E2 strength,
N. Auerbach, D.C. Zheng, L. II Zamick, B.Alex Brown, Phys. Lett. B 304 (1993) 17-23.
%
\bibitem{Troltenier1996}
Correlations between the quadrupole deformation, $B (E2; 0_1 \rightarrow 2_1 )$ value, and total GT + strength,
D. Troltenier, J.P. Draayer, J.G. Hirsch, Nucl. Phys. A 601 (1996) 89-102.
%
\bibitem{Zelevinsky2017}
Nuclear Structure Features of Gamow-Teller Excitations,
Vladimir Zelevinsky, Naftali Auerbach, Bui Minh Loc, Phys. Rev. C 96 (2017)  044319.
%
\bibitem{Radha1997}
Gamow-Teller strength distributions in fp-shell nuclei,
P. B. Radha, D. J. Dean, S. E. Koonin, K. Langanke, and P. Vogel,
Phys. Rev. C 56 (1997) 3079.
%
\bibitem{Langanke1997}
Shell model Monte Carlo studies of N = Z pf-shell nuclei with pairing-plus-quadrupole Hamiltonian,
K. Langanke, R Vogel, Dao-Chen Zheng, Nucl. Phys, A 626 (1997) 735-750.
%
\bibitem{Poves1998}
 Pairing and the structure of the pf-shell N~Z nuclei,
 Alfredo Poves, Gabriel Martinez-Pinedo,
 Phys. Lett. B 430 (1998) 203-208.
%
\bibitem{Martinez1999}
Competition of isoscalar and isovector proton-neutron pairing in nuclei,
G. Martínez-Pinedo  K. Langanke, P. Vogel, Nucl. Phys. A 651 (1999) 379-393.
%
\bibitem{Kaneko2018}
 Isoscalar neutron-proton pairing and SU(4)-symmetry breaking in Gamow-Teller transitions,
K. Kaneko, Y. Sun, and T. Mizusaki, Phys. Rev. C 97 (2018) 054326.
%
\bibitem{Bai2013}
Role of T = 0 pairing in Gamow-Teller states in N = Z nuclei,
C.L. Bai, H. Sagawa , M. Sasano, T. Uesaka, K. Hagino, H.Q. Zhang, X.Z. Zhang , F.R. Xu,, Phys. Lett. B 719 (2013) 116-120.
%
\bibitem{Sagawa2016}
Isovector spin-singlet (T = 1, S = 0) and isoscalar spin-triplet (T = 0, S = 1) pairing interactions and spin-isospin response,
H Sagawa, C L Bai and G Colò, Phys. Scr. 91 (2016) 083011 Invited Comment.
%
\bibitem{Isacker2018}
Gamow-Teller transitions and neutron-proton-pair transfer reactions,
P. Van Isacker, A.O. Macchiavelli, Phys. Lett. B 780 (2018) 414-417.
%
\bibitem{Antoine}
E. Caurier, shell model code ANTOINE, IRES, Strasbourg 1989-2004.
%
\bibitem{Caurier1999b} Present Status of Shell Model Techniques,
E. Caurier, F. Nowacki, Acta Physica Polonica 30 (1999) 705.
%
\bibitem{Caurier2005}
The Shell Model as Unified View of Nuclear Structure,
E. Caurier, G. Martinez-Pinedo, F. Nowacki, A. Poves, A.P. Zuker, Rev. Mod. Phys. 77 (2005) 427-488.
%
\bibitem{exp} National Nuclear Data Center, https://www.nndc.bnl.gov.
%
\bibitem{Fujita2015} High-resolution study of Gamow-Teller excitations in the $^{42}$Ca$(^3$He,$)^{42}$Sc
 reaction and the observation of a low-energy super-Gamow-Teller state, Y. Fujita et. al. Phys. Rev. C 91  (2015) 064316.
%
\bibitem{Fujita2013} High-resolution study of $T_z=+2 \rightarrow +1$ Gamow-Teller transitions in the $^{44}$Ca$(^3$He,$)^{44}$Sc reaction,
Y. Fujita et. al. Phys. Rev. C 88  (2013) 014308.
%
\bibitem{Ganio2016} High-resolution study of Gamow-Teller transitions in the $^{48}$Ti$(^3$He,$)^{48}$V reaction,
E. Ganio\u glu et. al. Phys. Rev. C 93 (2016) 064326.
%
\end{thebibliography}
\end{document}